\documentclass[longauth]{aa}

\usepackage[switch]{lineno}
\usepackage{graphicx}
\usepackage{arydshln}
\usepackage{subcaption}
\usepackage{hyperref}
\hypersetup{colorlinks=true,linkcolor=blue,filecolor=blue,urlcolor=blue,citecolor=blue}
\usepackage{appendix}

\usepackage{txfonts}
\begin{document}

   \title{Multi-year characterisation of the broad-band emission from the intermittent extreme BL Lac 1ES~2344+514}
   \titlerunning{Multi-year study of 1ES~2344+514}
%
\author{\normalsize
H.~Abe\inst{1} \and
S.~Abe\inst{1} \and
V.~A.~Acciari\inst{2} \and
I.~Agudo\inst{3} \and
T.~Aniello\inst{4} \and
S.~Ansoldi\inst{5,42} \and
L.~A.~Antonelli\inst{4} \and
A.~Arbet Engels\inst{6}\thanks{E-mail: \href{mailto:contact.magic@mpp.mpg.de}{contact.magic@mpp.mpg.de}. Corresponding authors: Axel Arbet Engels, Habib Ahammad Mondal, Satoshi Fukami, Filippo D'Ammando} \and
C.~Arcaro\inst{7} \and
M.~Artero\inst{8}
% ...
}

\institute { Japanese MAGIC Group: Institute for Cosmic Ray Research (ICRR), The University of Tokyo, Kashiwa, 277-8582 Chiba, Japan
\and Instituto de Astrof\'isica de Canarias and Dpto. de  Astrof\'isica, Universidad de La Laguna, E-38200, La Laguna, Tenerife, Spain
\and Instituto de Astrof\'isica de Andaluc\'ia-CSIC, Glorieta de la Astronom\'ia s/n, 18008, Granada, Spain
\and National Institute for Astrophysics (INAF), I-00136 Rome, Italy
\and Universit\`a di Udine and INFN Trieste, I-33100 Udine, Italy
\and Max-Planck-Institut f\"ur Physik, D-80805 M\"unchen, Germany
\and Universit\`a di Padova and INFN, I-35131 Padova, Italy
\and Institut de F\'isica d'Altes Energies (IFAE), The Barcelona InsINAF - Istituto di Radioastronomia, Via Gobetti 101, I-40129 Bologna, Italytitute of Science and Technology (BIST), E-08193 Bellaterra (Barcelona), Spain
\and Technische Universit\"at Dortmund, D-44221 Dortmund, Germany
\and Croatian MAGIC Group: University of Zagreb, Faculty of Electrical Engineering and Computing (FER), 10000 Zagreb, Croatia
\and IPARCOS Institute and EMFTEL Department, Universidad Complutense de Madrid, E-28040 Madrid, Spain
\and Centro Brasileiro de Pesquisas F\'isicas (CBPF), 22290-180 URCA, Rio de Janeiro (RJ), Brazil
\and Centro de Investigaciones Energ\'eticas, Medioambientales y Tecnol\'ogicas, E-28040 Madrid, Spain
\and ETH Z\"urich, CH-8093 Z\"urich, Switzerland
\and Departament de F\'isica, and CERES-IEEC, Universitat Aut\`onoma de Barcelona, E-08193 Bellaterra, Spain
\and Universit\`a di Pisa and INFN Pisa, I-56126 Pisa, Italy
\and Universitat de Barcelona, ICCUB, IEEC-UB, E-08028 Barcelona, Spain
\and Department for Physics and Technology, University of Bergen, Norway
\and INFN MAGIC Group: INFN Sezione di Catania and Dipartimento di Fisica e Astronomia, University of Catania, I-95123 Catania, Italy
\and INFN MAGIC Group: INFN Sezione di Torino and Universit\`a degli Studi di Torino, I-10125 Torino, Italy
\and INFN MAGIC Group: INFN Sezione di Bari and Dipartimento Interateneo di Fisica dell'Universit\`a e del Politecnico di Bari, I-70125 Bari, Italy
\and Croatian MAGIC Group: University of Rijeka, Faculty of Physics, 51000 Rijeka, Croatia
\and University of Geneva, Chemin d'Ecogia 16, CH-1290 Versoix, Switzerland
\and Japanese MAGIC Group: Physics Program, Graduate School of Advanced Science and Engineering, Hiroshima University, 739-8526 Hiroshima, Japan
\and Armenian MAGIC Group: ICRANet-Armenia, 0019 Yerevan, Armenia
\and University of Lodz, Faculty of Physics and Applied Informatics, Department of Astrophysics, 90-236 Lodz, Poland
\and Croatian MAGIC Group: Josip Juraj Strossmayer University of Osijek, Department of Physics, 31000 Osijek, Croatia
\and Universit\"at W\"urzburg, D-97074 W\"urzburg, Germany
\and Finnish MAGIC Group: Finnish Centre for Astronomy with ESO, University of Turku, FI-20014 Turku, Finland
\and Japanese MAGIC Group: Department of Physics, Tokai University, Hiratsuka, 259-1292 Kanagawa, Japan
\and Universit\`a di Siena and INFN Pisa, I-53100 Siena, Italy
\and Saha Institute of Nuclear Physics, A CI of Homi Bhabha National Institute, Kolkata 700064, West Bengal, India
\and Inst. for Nucl. Research and Nucl. Energy, Bulgarian Academy of Sciences, BG-1784 Sofia, Bulgaria
\and Japanese MAGIC Group: Department of Physics, Yamagata University, Yamagata 990-8560, Japan
\and Finnish MAGIC Group: Space Physics and Astronomy Research Unit, University of Oulu, FI-90014 Oulu, Finland
\and Japanese MAGIC Group: Chiba University, ICEHAP, 263-8522 Chiba, Japan
\and Japanese MAGIC Group: Institute for Space-Earth Environmental Research and Kobayashi-Maskawa Institute for the Origin of Particles and the Universe, Nagoya University, 464-6801 Nagoya, Japan
\vfill\null
\and Japanese MAGIC Group: Department of Physics, Kyoto University, 606-8502 Kyoto, Japan
\and INFN MAGIC Group: INFN Sezione di Perugia, I-06123 Perugia, Italy
\and INFN MAGIC Group: INFN Roma Tor Vergata, I-00133 Roma, Italy
\and Japanese MAGIC Group: Department of Physics, Konan University, Kobe, Hyogo 658-8501, Japan
\and also at International Center for Relativistic Astrophysics (ICRA), Rome, Italy
% ...
}

   \date{Received 31 August 2023; Accepted 28 September 2023}

 
  \abstract
   {}
   { The BL Lac 1ES~2344+514 is known for temporary extreme properties (e.g., a shift of the synchrotron spectral energy distribution (SED) peak energy $\nu_{\rm synch,p}$ above $1$\,keV). While those extreme states were so far observed only during high flux levels, additional multi-year observing campaigns are required to achieve a coherent picture. Here, we report the longest investigation of the source from radio to very high energy (VHE) performed so far, focusing on a systematic characterisation of the intermittent extreme states. }
   {We organised a monitoring campaign covering a 3-year period from 2019 to 2021. More than ten instruments participated in the observations in order to cover the emission from radio to VHE. In particular, sensitive X-ray measurements by \textit{XMM-Newton}, \textit{NuSTAR}, and \textit{AstroSat} took place simultaneously with multi-hour MAGIC observations, providing an unprecedented constraint of the two SED components for this blazar.}
   {While our results confirm that 1ES~2344+514 typically exhibits $\nu_{\rm synch,p}>1$\,keV during elevated flux periods, we also find periods where the extreme state coincides with low flux activity. A strong spectral variability thus happens in the quiescent state, and is likely caused by an increase of the electron acceleration efficiency without a change in the electron injection luminosity. On the other hand, we also report a strong X-ray flare (among the brightest for 1ES~2344+514) without a significant shift of $\nu_{\rm synch,p}$. During this particular flare, the X-ray spectrum is among the softest of the campaign. It unveils complexity in the spectral evolution, where the common ``harder-when-brighter'' trend observed in BL Lacs is violated. By combining \textit{Swift}-XRT and \textit{Swift}-UVOT measurements during a low and hard X-ray state, we find an excess of the UV flux with respect to an extrapolation of the X-ray spectrum to lower energies. This ``UV excess'' implies that at least two regions contribute significantly to the infrared/optical/ultraviolet/X-ray emission. Using the simultaneous MAGIC, \textit{XMM-Newton}, \textit{NuSTAR}, and \textit{AstroSat} observations, we argue that a region possibly associated with the 10\,GHz radio core may explain such an excess. Finally, we investigate a VHE flare, showing an absence of simultaneous variability in the 0.3--2\,keV band. Using a time-dependent leptonic modelling, we show that this behaviour, in contradiction to single-zone scenarios, can instead be explained by a two-component model.}
   {}


   \maketitle
%

\section{Introduction}
\label{introduction_section}

1ES~2344+514 \citep[RA=23$^\text{h}$47$^{\prime}$04.837$^{\prime\prime}$, Dec=+51$^{\circ}$42$^{\prime}$17.878$^{\prime\prime}$, J2000,][]{2020yCat.1350....0G} is a nearby BL Lacertae (BL Lac) object located at a redshift of $z=0.044$ \citep{1996ApJS..104..251P}. It is a member of the blazar category --- an active galactic nucleus (AGN) whose relativistic plasma jet is aligned with the observer's line of sight \citep{2017SSRv..207....5R}. As typically observed in blazars, the spectral energy distribution (SED) displays two broad emission components. The low-energy component ranges from radio to X-rays, and the high-energy component is located in the gamma-ray band. The low-energy component peaks above $10^{15}$\,Hz, implying that it belongs to the subcategory of high-frequency BL Lac objects \citep[HBL;][]{1995ApJ...444..567P}. 1ES~2344+514 is one of the first extragalactic objects detected at very high energy (VHE; $E>100$\,GeV). The first VHE detection was achieved by the Whipple 10\,m telescope during a bright flare in 1996 with a peak flux of $\sim 60\%$ of the Crab Nebula flux above 350\,GeV \citep{1998ApJ...501..616C}. \citet{2017MNRAS.471.2117A} reported an average flux above 350\,GeV of $\sim 4\%$ that of the Crab Nebula between 2008 and 2015 in the absence of any flaring activity. In what follows, we consider such a flux level as representative of the quiescent VHE activity of 1ES~2344+514. We note however that the VHE flux can vary within a factor $\sim$2 down to daily timescales \citep{2007ApJ...662..892A, 2011ApJ...738..169A, 2017MNRAS.471.2117A}. Regarding the X-ray band, the 2-10\,keV flux lies around $10^{-11}$\,erg\,cm$^{-2}$\,s$^{-1}$ in quiescent states \citep{2011ApJ...738..169A, 2013A&A...556A..67A}.\par

\citet{2000MNRAS.317..743G} reported strong spectral variability of the X-ray spectrum on a timescale of 5\,ks during a flaring state: the peak energy of the low-energy SED component shifted by a factor of more than 30, and reached energies above 10\,keV. HBLs with a low-energy component peaking above 1\,keV are dubbed as extreme high-frequency BL Lacs \citep[EHBL;][]{2001A&A...371..512C, 2020NatAs...4..124B}. Recently, \citet{2020MNRAS.496.3912M} published a multiwavelength study of 1ES~2344+514 in a flaring period that happened in 2016. The source again temporarily behaved as an EHBL. The broad-band SED could be well modelled both with leptonic and hadronic scenarios. In leptonic models, the low-energy SED component originates from electron-synchrotron radiation, while the high-energy component is attributed to electron inverse-Compton (IC) scattering off the synchrotron photons. This model is commonly dubbed as the synchrotron self-Compton model \citep[SSC; see, e.g.,][]{1992ApJ...397L...5M}. In the hadronic model, the low-energy component still originates from electron-synchrotron radiation, but the high-energy component is ascribed to proton-synchrotron radiation \citep[see, e.g.,][]{2001APh....15..121M, 2015MNRAS.448..910C}. During the quiescent activity, the peak energy of 1ES~2344+514 was estimated to be around 0.1\,keV \citep{2013A&A...556A..67A, 2018A&A...620A.185N, 2020ApJ...892..105A}. 1ES~2344+514 is thus characterised by an EHBL behaviour occurring on a temporary basis, which seems to happen mostly during high emission states. \par 

As highlighted by \citet{2020NatAs...4..124B}, the EHBL population is not homogeneous. While some of the members are EHBL-like only temporarily (e.g., 1ES~2344+514), other EHBLs display extreme properties on a constant basis (e.g., 1ES~1426+428). In addition to that, several EHBLs are not only extreme in the synchrotron domain but also in the VHE band with a high-energy SED component peaking above 1\,TeV (e.g., 1ES~0229+200). Those sources are commonly called \textit{extreme-TeV} EHBL and are particularly challenging for standard blazar acceleration and emission models \citep{2011A&A...534A.130K}. In general, EHBLs represent the most energetic class of blazars and their study is particularly relevant in the context of particle acceleration mechanisms in AGN jets.\par

The intermittent EHBL nature of 1ES~2344+514 is still poorly characterised owing to the low amount of multi-year broad-band campaigns performed so far. This prevents a detailed understanding of the physical origin of these extreme states and how exactly the EHBL state correlates with the flux activity. This work presents the longest multi-year study from radio to VHE of 1ES~2344+514 performed so far. A dense multiwavelength campaign was organised between 2019 and 2021 by involving more than ten different instruments. We emphasize that most of the observing campaign was organized through an unbiased monitoring of the source, without triggering observations on particular flaring events, in order to get a systematic investigation of the spectral evolution. A detailed characterisation of the X-ray emission was obtained using \textit{XMM-Newton}, Nuclear Spectroscopic Telescope Array (\textit{NuSTAR}), and \textit{AstroSat} deep exposures. The latter observations are accompanied by multi-hour MAGIC exposures with the aim to acquire a precise determination of the two SED components. Furthermore, thanks to a dense monitoring from the \textit{Neil Gehrels Swift Observatory} (\textit{Swift}) and MAGIC telescopes, we carry out a systematic investigation of the intermittent EHBL state in the synchrotron and VHE gamma-ray regimes. This paper also discusses an intriguing flare that was detected in 2019, during which the emission characteristics suggest at least two separate emitting components contributing to the SED from infrared (IR) to X-rays. Finally, the results are interpreted within theoretical leptonic models, which are able to successfully describe the SEDs.\par 

The paper is structured in the following way. Sect.~\ref{sec:analysis} describes the observations and data reduction, Sect.~\ref{sec:mwl_variability_sec} discusses the multiwavelength variability and the observed correlation trends over the campaign, and Sect.~\ref{sec:spectral_study} presents the spectral analysis in the X-ray \& VHE bands. The theoretical modelling of the deep exposures with simultaneous MAGIC, \textit{NuSTAR}, \textit{XMM-Newton}, and \textit{AstroSat} data is shown in Sect.~\ref{low_state_model}, a general discussion of the different results is presented in Sect.~\ref{sec:discussion}, and the conclusions are drawn in Sect.~\ref{sec:conclusion}.

\section{Dataset and analysis}
\label{sec:analysis}

\begin{table*}[h!]
\caption{\label{tab:XMM_nustar_exposures} Table summarizing the observing times of the VHE and X-ray instruments during the two long exposures in July 2020 and August 2021.} 
\setlength{\tabcolsep}{6pt} 
\renewcommand{\arraystretch}{1.3} 
\centering
\begin{tabular}{lccc}     
\hline\hline
 Epoch label & Instrument & Start \& End time [UTC] & Exposure [ks]\\
\hline
 & MAGIC & 2020-07-22T23:36:43 - 2020-07-23T04:49:13 & 17\\
Deep exposure 1 & \textit{XMM-Newton} & 2020-07-22T21:59:37 - 2020-07-23T06:01:17 & 19 (pn), 27 (MOS1/2) \\
 & \textit{NuSTAR} & 2020-07-22T19:26:10 - 2020-07-23T07:11:10 & 21 \\
 
\hdashline

 & MAGIC & 2021-08-05T23:32:52 - 2021-08-06T04:25:19 & 16\\
 Deep exposure 2 & \textit{XMM-Newton} & 2021-08-05T19:50:32 - 2021-08-06T04:25:32 & 19 (pn), 27 (MOS1/2)\\
 & \textit{AstroSat-SXT} & 2021-08-05T15:09:14 - 2021-08-07T01:21:42 & 41\\

\hline
\end{tabular}
\tablefoot{Regarding \textit{XMM-Newton}, the exposures are quoted separately for the EPIC pn and MOS1/MOS2 cameras.}

\end{table*}

\subsection{MAGIC}
\label{sec:magic_analysis}
The VHE observations were performed by the Florian Goebel Major Atmospheric Gamma Imaging Cherenkov telescopes (MAGIC) array, which consists of two 17\,m diameter imaging atmospheric Cherenkov telescopes located at an altitude of 2231\,m above the sea level, on the Canary Island of La Palma at the Roque de los Muchachos Observatory. The integral sensitivity of MAGIC for point-source observations above 220\,GeV is ($0.66\pm0.03$)\% of the Crab Nebula flux\footnote{For a given energy threshold, the Crab Nebula flux is defined as the integral flux of the Crab Nebula spectrum above the threshold energy. Throughout this work, we consider the Crab Nebula spectrum from \citet{2016APh....72...76A}.} in
50\,hr \citep{2016APh....72...76A}.\par


The MAGIC observations of 1ES~2344+514 covered a $\sim 2.5$\,yr period from August 2019 until December 2021. After data-quality selection, a total of about 32\,hr of effective observation time was collected in 25 nights. The source was observed with zenith angles ranging from $20^\circ$ to $62^\circ$. We use the standard analysis tools from the MAGIC Analysis and Reconstruction Software \citep[MARS; ][]{zanin2013, 2016APh....72...76A} to process the data. A fraction of the observations are affected by the presence of the Moon, which leads to an increased night-sky background light contamination \citep{2017APh....94...29A}. In order to take into account these varying observational conditions, the data are first split into several subsets according to the level of moonlight contamination. Then, the analysis is carried out by adopting Monte Carlo simulations tuned to match the corresponding conditions of the different data subsets \citep{2017APh....94...29A}.\par 

The MAGIC fluxes are computed above 300\,GeV in daily and yearly binning. They are plotted in Fig.~\ref{MWL_lightcurve}. The typical exposure of a single MAGIC observation lies between 0.5\,hr and 2\,hr. However, two nights have a significantly larger exposure of $\sim 5$\,hr, July 23\textsuperscript{rd} 2020 (MJD~59053) and August 6\textsuperscript{th} 2021 (MJD~59432). In the following, they are referred to as ``deep exposure 1'' and ``deep exposure 2''. They are marked in Fig.~\ref{MWL_lightcurve} with orange and brown dashed vertical lines, respectively. Those deep exposures took place simultaneously with sensitive X-ray observations by \textit{XMM-Newton}, \textit{NuSTAR}, and \textit{AstroSat} with the aim of obtaining precise simultaneous measurements of the low- and high-energy SED components of 1ES~2344+514. Table~\ref{tab:XMM_nustar_exposures} lists the corresponding exact observing times of MAGIC as well as of the accompanying X-ray instruments. Regarding ``deep exposure 1'', the MAGIC detection is significant and about $5.6\sigma$ (the measured flux is about 4\% of the Crab Nebula). As for the ``deep exposure 2'' epoch, the flux is lower (2\% of the Crab Nebula above 300\,GeV), leading to a detection significance of only $2\sigma$. No significant intranight VHE variability is detected in the MAGIC data on these two dates. The corresponding fluxes and spectra are thus averaged over their respective observation time. In this work, all MAGIC spectra and best-fit parameters are computed using a forward-folding method to take into account the finite energy resolution of the instrument \citep{zanin2013}. 

\subsection{\textit{Fermi}-LAT}

The Large Area Telescope (LAT) instrument is a pair-conversion telescope onboard the \textit{Fermi} satellite \citep{2009ApJ...697.1071A,2012ApJS..203....4A}. \textit{Fermi}-LAT surveys the gamma-ray sky in the 20\,MeV to $E>300$\,GeV energy range with an all-sky coverage on a $\sim 3$\,hr timescale. The analysis for this work is performed using an unbinned-likelihood approach with tools from the \texttt{FERMITOOLS} software\footnote{\url{https://fermi.gsfc.nasa.gov/ssc/data/analysis/}} v2.0.8isert. We adopt the instrument response function \texttt{P8R3\_SOURCE\_V2} and the diffuse background models\footnote{\url{http://fermi.gsfc.nasa.gov/ssc/data/access/lat/\\BackgroundModels.html}} \texttt{gll\_iem\_v07} and \texttt{iso\_P8R3\_SOURCE\_V3\_v1}.\par 

We select \texttt{Source} class events between 0.3\,GeV and 300\,GeV in a circular region of interest (ROI) with a radius of $15^\circ$ around 1ES~2344+514. The events with a zenith angle $>90^\circ$ are discarded to limit the contribution from gamma rays grazing Earth’s limb. To model the field sources, we consider all sources from the fourth {\it Fermi}-LAT source catalogue Data Release 2 \citep[4FGL-DR2;][]{2020ApJS..247...33A, 2020arXiv200511208B} that are located within the ROI plus an annulus of $5^\circ$. 1ES~2344+514 is modelled using a simple power-law function. In order to build light curves, the source model is fitted to the data by setting the normalisation and the spectral index of all the sources within 7$^\circ$ from the target as free parameters. Above $7^\circ$ all parameters are fixed to the 4FGL-DR2 values. The normalisations of the background components are left as free parameters. When the fit does not converge, the model parameters are fixed to the 4FGL-DR2 values for sources detected with a test statistic \citep[TS;][]{1996ApJ...461..396M} below 4. If after that the fit still does not converge, we gradually increase the TS threshold below which the model parameters are fixed, until a convergence is achieved. Additionally, the power-law index of 1ES~2344+514 is fixed to the 4FGL-DR2 value if the source is detected with TS $< 15$. Finally, for each time bin resulting in TS $< 5$, a flux upper limit at 95\% confidence level is quoted in the light curve.\par  

The light curve is computed in monthly time bins, which is the exposure time needed by \textit{Fermi}-LAT to detect 1ES~2344+514 during quiescent emission states ($F_{0.3-300\textrm{\,GeV}}\approx0.5\times 10^{-8}$\,cm$^{-2}$s$^{-1}$). The light curve is computed in 2-day bins close to a VHE flare detected in August 2019 (see Sect.\ref{sec:mwl_variability_sec}).

\subsection{\textit{Swift}-XRT}

To accompany the MAGIC monitoring, we organised many simultaneous X-ray observations by the \textit{Swift} X-ray Telescope \citep[XRT;][]{2005SSRv..120..165B}. The \textit{Swift}-XRT observations were performed in the Windowed Timing (WT) and Photon Counting (PC) readout modes depending on the source flux. The data are processed using the XRTDAS software package (v.3.7.0) developed by the ASI Space Science Data Center\footnote{\url{https://www.ssdc.asi.it/}} (SSDC), released by the NASA High Energy Astrophysics Archive Research Center (HEASARC) in the HEASoft package (v.6.30.1). The calibration files from \textit{Swift}-XRT CALDB (version 20210915) are used within the \texttt{xrtpipeline} to calibrate and clean the events.\par

The X-ray spectrum from each observation is extracted from the calibrated and cleaned event file. For both WT and PC modes data, the events for the spectral analysis is selected within a circle of 20 pixel ($\sim 47''$) radius. The background is extracted from a nearby circular region with the same radius. The ancillary response files are generated with the \texttt{xrtmkarf} task applying corrections for point spread function (PSF) losses and CCD defects using the cumulative exposure map.\par 

The 0.3--10\,keV source spectra are binned using the \texttt{grppha} task to ensure a minimum of 20 counts per bin, and then modelled in \texttt{XSPEC} using power-law and log-parabola models (with a pivot energy fixed at 1\,keV). Additionally, and for each X-ray analysis presented in this work, a photoelectric absorption component is included considering a column density fixed to $N_{\rm H}=1.41\times10^{21}$\,cm$^{-2}$ \citep[][]{2016A&A...594A.116H}. In the vast majority of the observations, the statistical preference for a log-parabola model is not significant. We compute fluxes in the energy bands
0.3--2\,keV and 2--10\,keV.

\subsection{\textit{XMM-Newton} EPIC}

We organized two X-ray observations of 1ES~2344+514 from \textit{XMM-Newton} \citep{2001A&A...365L...1J} during the MAGIC multi-hour observations on July 23\textsuperscript{rd} 2020 (``deep exposure 1'') and August 6\textsuperscript{th} 2021 (``deep exposure 2''). Table~\ref{tab:XMM_nustar_exposures} summarizes the exact observing windows as well as the duration. All three EPIC cameras (pn, MOS1, and MOS2) were operated in Large Window mode. The data are reduced using the \textit{XMM-Newton} Science Analysis System (SAS v20.0.0) following standard procedures. Time intervals with strong background flaring are filtered out following standard procedures using the high-energy light curves with cuts of 0.4 and 0.35 counts s$^{-1}$ for the pn and MOS, respectively. The total good exposure times after the filtering are 18.9, 26.7, and 26.7\,ks in 2020 and 19.4, 27.4, and 27.3\,ks in 2021 for the pn, MOS1, and MOS2, respectively. Source and background spectra are extracted from circular regions of radius $34''$ for all three detectors. All spectra are binned to contain at least 20 counts per bin and not to oversample the intrinsic energy resolution by more than a factor of three. The empirical correction of the EPIC effective area based on simultaneous {\em NuSTAR} observations is applied by setting applyabsfluxcorr=yes
in the \texttt{arfgen} task\footnote{\url{https://xmmweb.esac.esa.int/docs/documents/CAL-TN-0230-1-3.pdf}}. The spectra are finally modelled with a log-parabola function in the 0.4--10\,keV range using \texttt{XSPEC}  (with a pivot energy fixed at 1\,keV).\par

\subsection{\textit{NuSTAR}}

{\em NuSTAR} \citep[][]{harrison13} observed 1ES~2344+514 in the 3--79\,keV band during the MAGIC long exposure on July 23\textsuperscript{rd} 2020 (``deep exposure 1'')  with its two coaligned X-ray telescopes and corresponding focal planes, focal plane module A (FPMA) and B (FPMB). The total observing time is 21\,ks (see Table~\ref{tab:XMM_nustar_exposures}). The Level-1 data products are processed with the {\em NuSTAR} Data Analysis Software (\texttt{nustardas}, v1.9.7) within the HEAsoft software. Cleaned event files (Level-2 data products) are produced and calibrated using standard filtering criteria with the \texttt{nupipeline} software module, and the OPTIMIZED parameter for the exclusion of the South Atlantic Anomaly (SAA) passages. We use the calibration files available in the {\em NuSTAR} CALDB version 20220510. 

Spectra of the source are extracted for the whole observation from the cleaned event files using a circle of 26 pixel ($\sim 60''$) radius, while the background is extracted from a nearby circular region of 26 pixel radius on the same chip of the source. The choice of the extraction region size optimizes the signal-to-noise ratio, but alternative choices do not affect the results. The ancillary response files are generated with the \texttt{numkarf} task, applying corrections for the PSF losses, exposure maps, and vignetting. The spectra are rebinned with a minimum of 20 counts per energy bin to allow for $\chi^{2}$ spectrum fitting.

\subsection{\textit{AstroSat}}
Between August 5\textsuperscript{th} 2021 and August 7\textsuperscript{th} 2021, we obtained a deep exposure from the Soft X-ray Telescope (SXT), which is an X-ray imaging instrument onboard the \textit{AstroSat} satellite \citep{kpsing16,kpsing17}. The observation encompasses the one from MAGIC and \textit{XMM-Newton} during the ``deep exposure 2'' epoch.\par 


Level-1 data are stored, in FITS format, in the \textit{AstroSat} data archive\footnote{\url{https://astrobrowse.issdc.gov.in/astro_archive/archive/Home.jsp}}. The Level-2 data are generated from the Level-1 data by running the \texttt{sxtpipeline} tool provided by the SXT data-analysis package (\textit{AS1SXTLevel2-1.4b}). After merging the cleaned event files of individual orbits together by \texttt{sxtpyjuliamerger} we use the \texttt{xselect} tool of HEASoft to extract the images, light curves and spectra from the merged clean event files in the range 0.7--7\,keV. We select the source region as a circle of radius $12'$ centered at the point source. As background we use the file \textit{SkyBkg{\_}comb{\_}EL3p5{\_}Cl{\_}Rd16p0{\_}v01.pha} provided by SXT POC team. The fluxes (in the 0.7--7\,keV band) and spectral parameters were computed using \texttt{XSPEC} using a log-parabola model. Table~\ref{tab:XMM_nustar_exposures} summarises the exact observing window and exposure of \textit{AstroSat-SXT}.


\subsection{\textit{Swift}-UVOT}

We obtained ultraviolet (UV) data coverage with the \textit{Swift} UV/Optical Telescope \citep[UVOT;][]{2005SSRv..120...95R} observations between January 2019 and December 2021 performed in the $W1$, $M2$, and $W2$ filters. Over this time range, a total of 67 observations of 1ES~2344+514 are selected after quality checks. We perform photometry over the total exposures of each observation in the sample extracted from the official archive, applying the same apertures for source counts (the standard with $5''$ radius) and background (mostly three or four circles of $\sim 16''$ radii off the source) estimation. The photometry is obtained executing the task within the official software included in the HEAsoft 6.23 package, from HEASARC, and then applying the official calibrations \citep{2011AIPC.1358..373B} included in the more recent CALDB release (20201026). The fluxes are dereddened considering a mean interstellar extinction curve taken from \citet{1999PASP..111...63F} and a Galactic $E(B-V)$ value of 0.1805\,mag \citep{1998ApJ...500..525S, 2011ApJ...737..103S}.

\subsection{\textit{XMM-Newton} OM}

In addition to the X-ray observations from the three \textit{XMM-Newton} EPIC cameras, we process the data collected simultaneously by the Optical Monitor (OM) in the $B$, $U$, $W1$, $M2$, and $W2$ filters. The data reduction is performed using the SAS task \texttt{omichain}. The transformation from count rate to flux is achieved by using the conversion factors given in the SAS
watchout dedicated page\footnote{\url{https://www.cosmos.esa.int/web/xmm-newton/sas-watchout-uvflux}}. Magnitudes are finally corrected for Galactic extinction using an $E(B-V)$ value of 0.1805\,mag \citep{2011ApJ...737..103S}.

\subsection{Optical}

At optical wavelengths, we profited from $R$-band observations carried out by the Tuorla blazar monitoring program\footnote{\url{http://users.utu.fi/kani/1m/}} between January 3\textsuperscript{rd} 2019 (MJD~58486) and November 30\textsuperscript{th} 2021 (MJD~59458). The data were acquired by the Kungliga Vetenskapsakademien (KVA; 35\,cm primary mirror diameter) telescope and Joan Oró Telescope (TJO; 80\,cm primary mirror diameter). The KVA is located on the island of La Palma, in Spain, at the Roque de los Muchachos Observatory, while the TJO telescope is placed at the Montsec Astronomical Observatory, also in Spain. Data reduction is performed following the differential photometry method described by \citet{2018A&A...620A.185N} with an aperture radius of $7.5''$. The contributions of the host galaxy and nearby companions are subtracted from the observed fluxes, and we apply a correction for the Galactic extinction.\par

Additional $R$-band observations were performed by the Whole Earth Blazar Telescope\footnote{\url{http://www.oato.inaf.it/blazars/webt/}} \citep[WEBT; e.g.,][]{2007A&A...464L...5V,2017Natur.552..374R} consortium. The observations were made within the GLAST-AGILE Support Program \citep[GASP; e.g.,][]{2009A&A...504L...9V}, which provides mainly optical, but also radio and near-IR support to blazar observations by gamma-ray satellites.  Optical data for this paper are taken at the Abastumani, Athens, Crimean\footnote{In 1991, Ukraine with the Crimean peninsula became an
independent state. While the Crimean Astrophysical Observatory
became Ukrainian, the AZT-8 telescope located there continued
to be operated jointly by the Crimean Observatory and by the
St. Petersburg group.}, Haleakala, Lulin, McDonald, Perkins, Rozhen, Skinakas, St. Petersburg, Teide, Tijarafe, and Vidojevica Observatories. The $R$-band flux densities of the source are corrected for a quantity accounting for the contribution by the host galaxy and nearby companions, the Galactic extinction, and intercalibration among the different datasets. For the latter, we use the data by the Tuorla blazar monitoring program as reference.



Finally, we also make use of $R$-band observations from the 0.76\,m Katzman Automatic Imaging Telescope \citep[KAIT;][]{2001ASPC..246..121F} at Lick Observatory on Mt. Hamilton, CA, USA. The data are first reduced adopting the custom pipeline presented by \citet{2010ApJS..190..418G}. Then, the photometry is carried out using a 9-pixel aperture (corresponding to $7.2''$). Several nearby stars are chosen from the Pan-STARRS1\footnote{\url{http://archive.stsci.edu/panstarrs/search.php}} catalog for calibration, and their magnitudes are transformed into the \\citet{1983AJ.....88..439L, 1990PASP..102.1382L} system using the empirical prescription presented by \citet{2012ApJ...750...99T}, Eq.~6. Data from KAIT are corrected for Galactic extinction, and the contribution of the host galaxy (plus nearby companions) is subtracted with the procedure described above.\par

\subsection{OVRO}
The radio observations were performed by the Owens Valley Radio Observatory (OVRO) 40\,m telescope within the blazar monitoring program \citep{2011ApJS..194...29R}. OVRO employs an off-axis dual-beam optics and a cryogenic pseudo-correlation receiver with
a 15\,GHz center frequency and 3\,GHz bandwidth. The calibration is done using a temperature-stable diode noise source in order to remove receiver gain drifts. Finally, the flux-density scale is derived from observations of 3C~286 assuming a value of 3.44\,Jy at 15.0\,GHz from \citet{1977A&A....61...99B}. The flux-density scale has a systematic uncertainty of $\sim 5\%$, which is not included in the error bars of data points shown later. More details about the OVRO data reduction and calibration are provided by \citet{2011ApJS..194...29R}.

\section{Multiwavelength light curves}
\label{sec:mwl_variability_sec}

\begin{figure*}
   \centering
   \includegraphics[width=2\columnwidth]{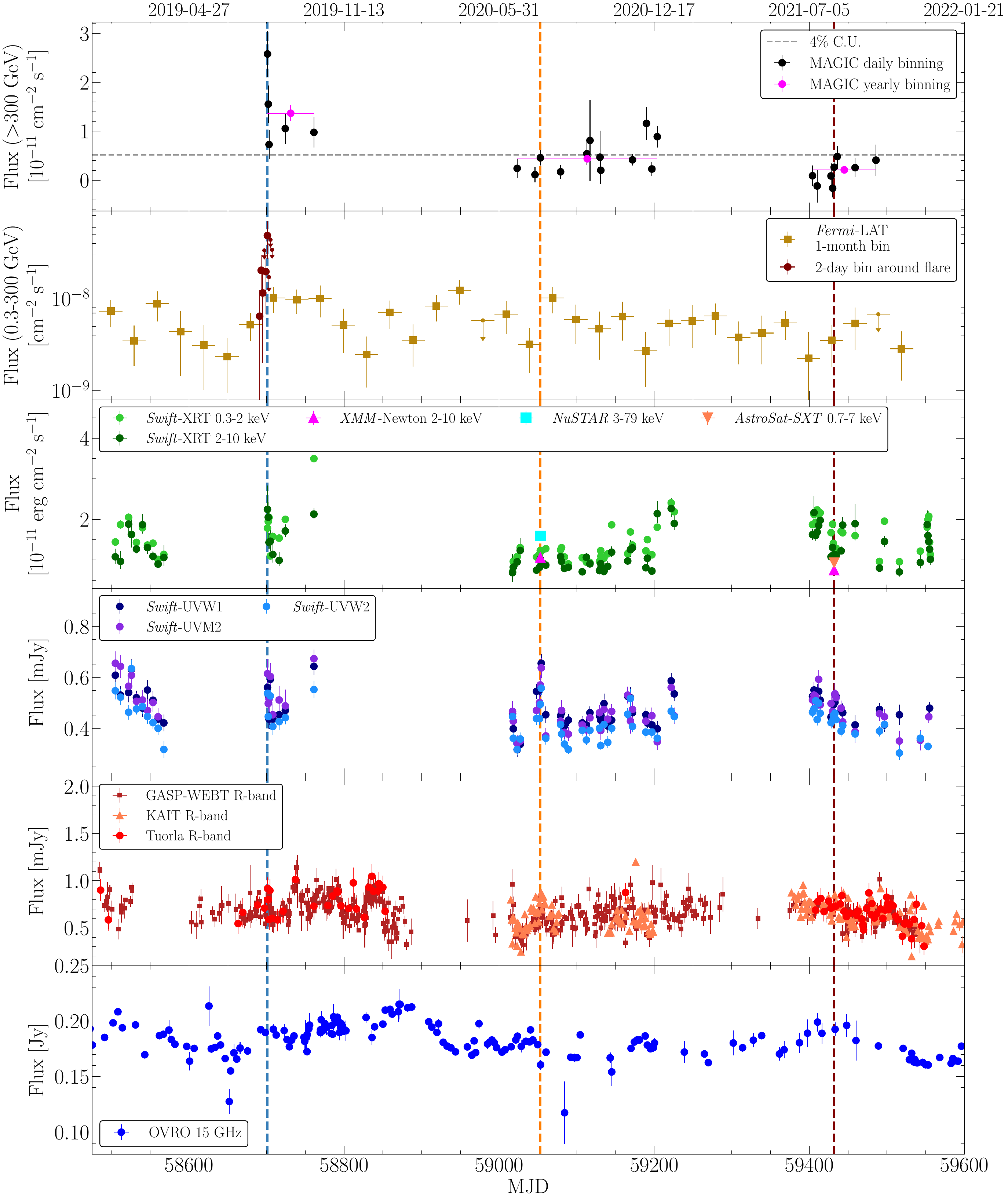}
   \caption{\tiny Multiwavelength light curves between December 23\textsuperscript{rd} 2018 (MJD~58475) and January 21\textsuperscript{st} 2022 (MJD~59600). The vertical dashed blue line highlights the date of the VHE flare detected by MAGIC in August 2019. The orange and maroon dashed vertical lines highlight the dates of the ``deep exposure 1'' (July 23\textsuperscript{rd} 2020 - MJD~59053) and ``deep exposure 2'' (August 6\textsuperscript{th} 2021 -- MJD~59432) epochs that contain simultaneous long observations from MAGIC, \textit{XMM-Newton}, \textit{NuSTAR}, and \textit{AstroSat}. The top panel shows the MAGIC fluxes above 300\,GeV in daily (black markers) and yearly binning (pink markers). The horizontal dashed-grey line represents 4\% of the Crab Nebula flux above 300\,GeV. The second panel from the top reports the \textit{Fermi}-LAT fluxes in the 0.3--300\,GeV band in monthly binning. An upper limit at 95\% confidence level is quoted for time bins with TS $<5$. The maroon markers are fluxes in 2-day binning around the VHE flare. In the third panel from the top, the \textit{Swift}-XRT fluxes are shown in the 0.3--2\,keV (light-green markers) and 2--10\,keV (dark-green markers) bands. The \textit{NuSTAR} (cyan), \textit{XMM-Newton} (pink), and \textit{AstroSat-SXT}2022). The gray bands correspond to the three (light red) fluxes are shown in the 3--79\,keV, 2--10\,keV, and 0.7--7\,keV bands, respectively. The fourth panel from the top displays the UV fluxes from the \textit{Swift}-UVOT instrument in the $UVW1$, $UVM2$, and $UVW2$ filters. The fifth panel from the top reports the fluxes in the $R$ band from GASP-WEBT, KAIT, and Tuorla. Finally, the bottom panel shows the OVRO fluxes measured at 15\,GHz.}
    \label{MWL_lightcurve}
\end{figure*}

\begin{figure}[h!]
   \centering
   \includegraphics[width=\columnwidth]{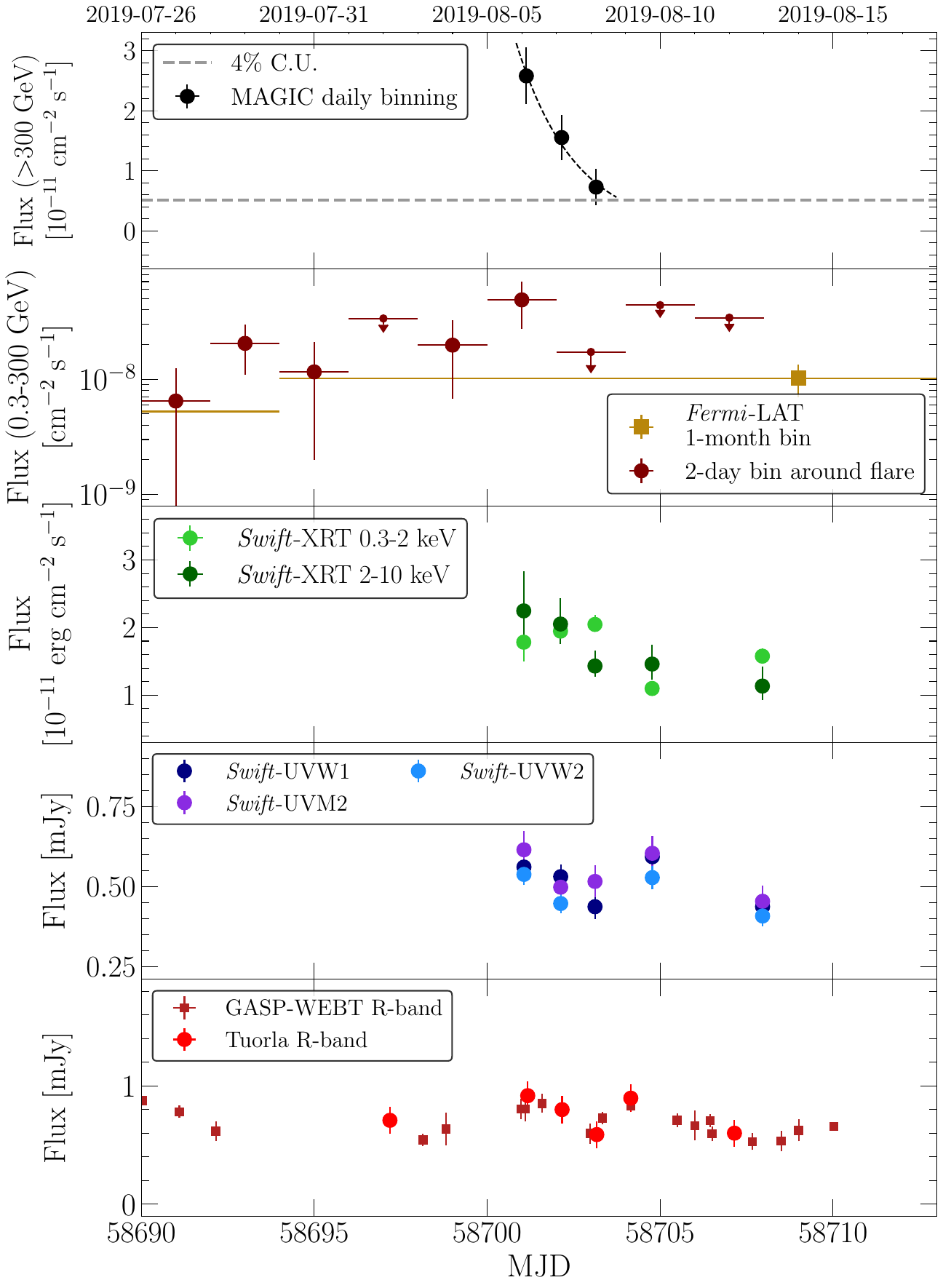}
   \caption{\small Multiwavelength light curves around the VHE flare detected by MAGIC in August 2019. The top panel shows the MAGIC fluxes above 300\,GeV in daily binning. The dashed curve is a fit of an exponential function using Eq.~\ref{decay_eq} in order to estimate the variability timescale (see text for more details). The horizontal dashed-grey line represents 4\% of the Crab Nebula flux above 300\,GeV. The second panel from the top displays the \textit{Fermi}-LAT fluxes in the 0.3--300\,GeV band in monthly binning (dark-yellow markers) and 2-day binning (maroon markers). In case the source is detected with TS $<5$, an upper limit at 95\% confidence level is quoted. In the third panel from the top are the \textit{Swift}-XRT fluxes in the 0.3--2\,keV (light-green markers) and 2--10\,keV (dark-green markers) bands. The fourth panel from the top displays the UV fluxes from the \textit{Swift}-UVOT instrument in the $UVW1$, $UVM2$, and $UVW2$ filters. Finally, the bottom panel shows the $R$-band fluxes from GASP-WEBT and Tuorla.}
    \label{MWL_lightcurve_flare}
\end{figure}

In Fig.~\ref{MWL_lightcurve} the multiwavelength light curves between December 23\textsuperscript{rd} 2018 (MJD~58475) and January 21\textsuperscript{st} 2022 (MJD~59600) are displayed from radio to VHE. In the top panel, the MAGIC fluxes above 300\,GeV are plotted with daily binning (dark markers) and yearly binning (pink markers). The horizontal grey dashed line depicts 4\% of the Crab Nebula flux, which is a good approximation of the quiescent state of 1ES~2344+514 at VHE. It corresponds to the average flux level reported by \citet{2017MNRAS.471.2117A} between 2008 and 2015 when no VHE flaring activity was detected. A VHE flare is observed in August 2019 and is highlighted with a vertical blue dashed line. This flare is studied in greater detail in Sect.~\ref{sec:flare_LC} and Sect.~\ref{flare_modelling}. During the rest of the campaign, no strong flare is detected at VHE. The 2020 average flux above 300\,GeV is $(3.5\pm0.4)\%$ that of the Crab Nebula, in agreement with the quiescent state. For 2021, the average flux drops to $(1.9\pm0.6)\%$ of that of the Crab Nebula and is one of the lowest states measured for 1ES~2344+514.\par

In the second panel from the top, the \textit{Fermi}-LAT fluxes in the 0.3--300\,GeV band are shown with monthly binning in dark-yellow markers. Throughout the campaign the monthly fluxes fluctuate around the quiescent state of the source. The comparison of the 2019--2021 emission with the long-term behaviour can be seen in Fig.~\ref{fermi_radio_lc} of Appendix~\ref{radio_fermi_longterm_lc}, which presents a \textit{Fermi}-LAT light curve starting from 2008. Close to the VHE flare in August 2019, the monthly emission shows little variability. However, a 2-day binned light curve simultaneous with the flare (maroon markers in Fig.~\ref{MWL_lightcurve}) indicates a clear flux increase on shorter timescale (see also Sect.~\ref{sec:flare_LC}).\par 

In the X-ray, the \textit{Swift}-XRT fluxes in the 0.3--2\,keV and 2--10\,keV bands are binned observation-wise (with a typical exposure time around 1--2\,ks) and show variability on a daily timescale. The 2--10\,keV flux varies around $1\times 10^{-11}$--$2\times 10^{-11}$\,erg\,cm$^{-2}$\,s$^{-1}$,  which is the typical dynamical range for 1ES~2344+514 in quiescent activity \citep{2011ApJ...738..169A, 2013A&A...556A..67A}. Nonetheless, a bright flare in the 0.3--2\,keV band is visible on October 5\textsuperscript{th} 2019 (MJD~58761), while the 2--10\,keV flux remains at the quiescent state. This particular night appears as an outlier with respect to the other nights and is discussed in more detail in Sect.~\ref{sec:X-ray_spectral_study}.\par   

The two long X-ray exposures accompanying the multi-hour MAGIC observations are highlighted in Fig.~\ref{MWL_lightcurve} with vertical orange and maroon dashed lines. The first one, labelled as ``deep exposure 1'' (on July 23\textsuperscript{rd} 2020), includes simultaneous \textit{XMM-Newton} and \textit{NuSTAR} pointings, and the second one, labelled as ``deep exposure 2'' (on August 6\textsuperscript{th} 2021) comprises simultaneous \textit{XMM-Newton} and \textit{AstroSat} pointings (see Table~\ref{tab:XMM_nustar_exposures}). The \textit{XMM-Newton} and \textit{NuSTAR} fluxes are plotted in pink and cyan colors, respectively. During these two epochs the VHE flux source is low: about 4\% of the Crab Nebula for the ``deep exposure 1'' epoch and about 2\% of the Crab Nebula regarding ``deep exposure 2''.\par 

The organisation of multiband long exposures represents the only possibility to search for flux variations down to sub-hour scales, being the timescale over which blazars are known to vary. Such investigations are also essential to provide constraints on the source dimension based on causality arguments. Our data reveal no strong variability at VHE nor in the X-ray during both ``deep exposure 1'' and ``deep exposure 2'' epochs. The MAGIC fluxes are fully consistent with a constant behaviour, while the X-ray emission (in \textit{XMM-Newton}, \textit{NuSTAR}, or \textit{AstroSat} data) varies at the level of 10\% only. Nonetheless, we exploit these observations to achieve a precise spectral characterisation of the low activity of the source, which is studied in detail in Sect.~\ref{sec:X-ray_spectral_study} and Sect.~\ref{low_state_model}.\par 

In the UV, optical, and radio wavebands, no apparent flaring episode is noted throughout the multiwavelength campaign. As discussed and quantified in Sect.~\ref{MWL_variability}, those energy regions of the spectrum show a smaller variability strength compared to the X-ray and VHE bands.\par 

Eventually, we search for correlated variability between the different bands between 2019 and 2021. Marginal hints of correlation are found between the VHE and X-ray, and the results are shown in Sect.~\ref{sec:X-ray_vhe_correlation_study}. No significant correlation is found between the other bands over the multiwavelength campaign presented in this work. In Sect.~\ref{sec:radio_he_correlation_study} we search for correlation between the \textit{Fermi}-LAT and OVRO fluxes using $\sim 13$\,yr of data, revealing a marginal hint.

\begin{figure}
   \centering
   \includegraphics[width=\columnwidth]{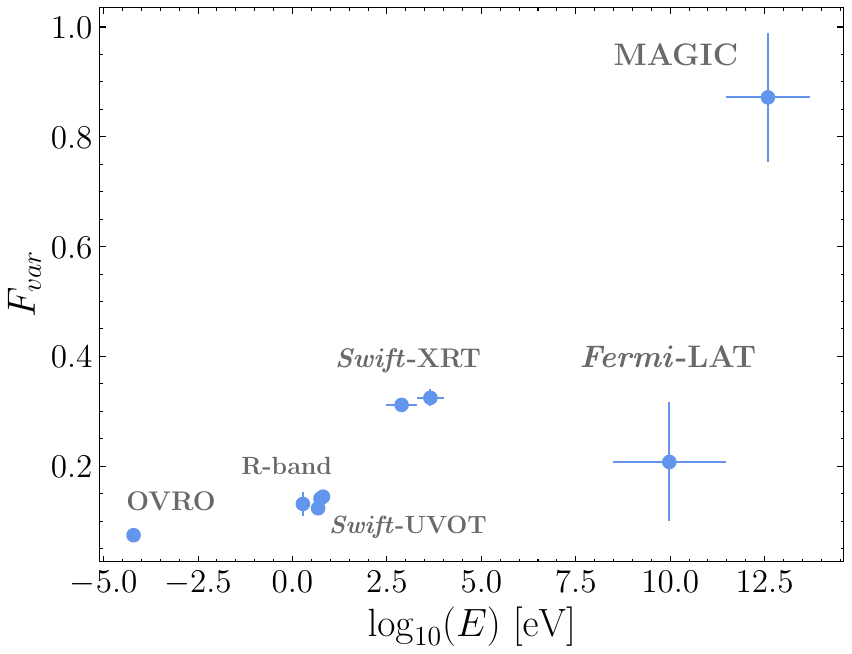}
   \caption{Fractional variability as a function of energy for the multiwavelength light curves in Fig.~\ref{MWL_lightcurve} . }
    \label{Fvar}
\end{figure}  

\subsection{Zoom on the August 2019 flare}
\label{sec:flare_LC}

Fig.~\ref{MWL_lightcurve_flare} is a zoom around the VHE flare detected with MAGIC in 2019. The top panel displays the MAGIC fluxes. The highest flux is measured on August 6\textsuperscript{th} 2019 (MJD~58701), and is about 20\% that of the Crab Nebula above 300\,GeV. This is $\gtrsim 5$ times larger than the quiescent state of 1ES~2344+514 at VHE \citep{2017MNRAS.471.2117A}. The subsequent daily measurements show a continuous flux dimming, which is inconsistent with a constant-flux hypothesis at the level of $3\sigma$. We estimate the decay timescale, $t_{\rm decay}$, by fitting an exponential function \citep[see e.g.][]{2010ApJ...722..520A}:

\begin{equation}
    F_{>300\mathrm{GeV}}(t)= F_0 \, e^{-(t-t_{\rm peak})/t_{\rm decay}}\, , 
    \label{decay_eq}
\end{equation}
\noindent
where $t_{\rm peak}$ is the time of the maximum flux that we fix to the center of the first MAGIC observation. Based on Eq.~\ref{decay_eq}, we find $t_{\rm decay} = 1.7 \pm 0.5$\,day. The result of the fit is shown as a black dotted line in the MAGIC panel of Fig.~\ref{MWL_lightcurve_flare}.\par 

The \textit{Fermi}-LAT fluxes in brown markers are binned in 2 days and also show a hint of a roughly daily timescale flare simultaneous with the VHE. The estimated flux for the time bin centered on August 6\textsuperscript{th} 2019 (the first MAGIC observation) is the highest and about 5 times that of the monthly average (dark-yellow markers). The hint of the flare is strengthened by the fact that on the date of the highest \textit{Fermi}-LAT flux, the source is significantly detected with TS $\approx 45$ ($\sim 7\sigma$) over a only 2-day integration time. We also find TS $< 18$ for the other 2-day bins around the VHE flare. In fact, a significant \textit{Fermi}-LAT detection (i.e., TS $\approx25$) of 1ES~2344+514 requires an integration time on roughly weekly timescale for its quiescent 0.3--300\,GeV flux of $\sim 0.5 \times10^{-8}$\,cm$^{-2}$s$^{-1}$.\par 

In the X-ray, the flux evolution is statistically consistent with a constant hypothesis within $2\sigma$ both in the 0.3--2\,keV and 2--10\,keV bands. A hint of flux decay is nevertheless apparent in the 2--10\,keV regime. Regarding the optical and UV, here also the flux is consistent with a constant evolution. As argued later in Sect.~\ref{sec:X-ray_vhe_correlation_study} and Sect.~\ref{flare_modelling}, this broad-band behaviour is difficult to reconcile with a single one-zone leptonic model and requires an additional emission component developing in the jet. A time-dependent modelling of the three consecutive days is performed in Sect.~\ref{flare_modelling}.

\begin{figure*}[h!]
   \includegraphics[width=2\columnwidth]{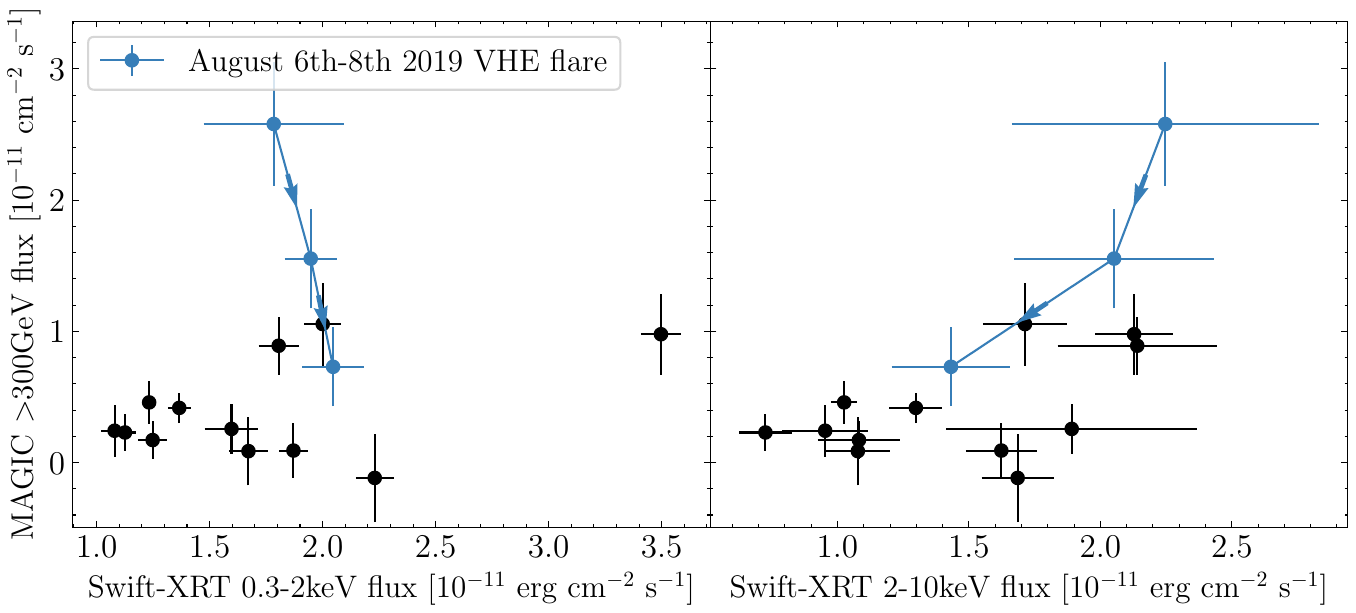}
   \caption{VHE flux versus X-ray flux over the multiwavelength campaign using MAGIC and \textit{Swift}-XRT data. Fluxes are nightly binned. Only pairs of measurements within 4\,hr are considered. The blue measurements correspond to the flaring state in August 2019 and the arrows show the direction of time. The Pearson coefficients and DCF computed over the full data sample} are shown in Table~\ref{tab:correlation_results_VHE_swift}.
    \label{VHE_vs_swift}
\end{figure*}
 
\begin{table}[h!]
\caption{\label{tab:correlation_results_VHE_swift} Results of the MAGIC versus \textit{Swift}-XRT correlations shown in Fig.~\ref{VHE_vs_swift}.} 
\setlength{\tabcolsep}{5pt} 
\renewcommand{\arraystretch}{2.} 
\centering
\begin{tabular}{l|cc}     
\hline\hline
 Energy bands & Pearson ($\sigma$) & DCF \\
\hline
$>300$\,GeV vs. 0.3--2\,keV& $0.2_{-0.3}^{+0.2}$ (0.9) & $0.3\pm0.2$ \\

\hdashline

$>300$\,GeV vs. 2--10\,keV & $0.6_{-0.2}^{+0.1}$ (2.7) & $0.7\pm0.4$ \\

\hline
\end{tabular}
\tablefoot{
The uncertainty and the significance (in $\sigma$ units) of the Pearson's coefficients are computed following \citet{2002nrca.book.....P}.
}
\end{table}

\subsection{Variability}
\label{MWL_variability}

\noindent The variability of the source throughout the spectrum is characterised using the fractional variability $F_{\rm var}$ \citep[][]{Vaughan2003} in different wavebands. The fractional variability is essentially the normalised variance of the flux after subtracting statistical fluctuations. It is defined as:
\begin{equation}
    F_{\rm var}=\sqrt{\frac{S^2- \langle \sigma_{err}^2 \rangle}{\langle x \rangle^2}} \, ,
\end{equation}
\noindent where $S$ is the standard deviation of the flux for $N$ measurements, $\langle x \rangle$ is the average flux and $\langle \sigma_{err}^2 \rangle$ is the corresponding mean square error. Following the prescription of \citet{Poutanen2008}, the uncertainty in $F_{\rm var}$ is estimated as:
\begin{equation}
    \Delta F_{\rm var} = \sqrt{F_{\rm var}^2 + err(\sigma_{NXS}^2)} - F_{\rm var} \, ,
\end{equation}
where,
\begin{equation}
    err(\sigma_{NXS}^2) = \sqrt{\left(\sqrt{\frac{2}{N}} \cdot \frac{\langle \sigma_{err}^2 \rangle }{\langle x \rangle ^2}\right)^2 + \left(\sqrt{\frac{\langle \sigma_{err}^2 \rangle }{N} } \cdot \frac{2F_{\rm var}}{\langle x \rangle}\right)^2} \, .
\end{equation}

The results are shown in Fig.~\ref{Fvar} and are obtained using the light curves displayed in Fig.~\ref{MWL_lightcurve}, meaning using all data between December 23\textsuperscript{rd} 2018 (MJD~58475) and January 21\textsuperscript{st} 2022 (MJD~59600).\par 

$F_{\rm var}$ shows a roughly monotonic increase from the radio to the X-ray (corresponding to the low-energy SED component of 1ES~2344+514) as well as from the MeV to TeV energies (corresponding to the high-energy SED component). The highest variability is found in the MAGIC fluxes. Such a two-peak structure of $F_{\rm var}(E)$ is typical in HBLs \citep[see e.g.][]{2015A&A...576A.126A} and the locations of the two local maxima match roughly the peak frequency of the two SED components, located in the X-ray (probed by \textit{Swift}-XRT) and VHE bands (probed by MAGIC), respectively. This behaviour of $F_{\rm var}(E)$ is attributed to the fact that towards higher energies in the respective SED components, the flux is radiated by particles with increasing energy (for instance, the characteristic synchrotron frequency is $\nu \propto B' \gamma'^2$, where $B'$ and $\gamma'$ are the magnetic field and electron Lorentz factor, respectively). More energetic particles suffer from stronger and faster cooling (from synchrotron and IC emission), hence leading to higher variability with increasing photon energy.

\subsection{VHE versus X-ray correlation}
\label{sec:X-ray_vhe_correlation_study}

We characterise the VHE versus X-ray correlation over the full campaign using the MAGIC and  \textit{Swift}-XRT observations. We correlate the MAGIC flux above 300\,GeV with the \textit{Swift}-XRT flux estimated in the 0.3--2\,keV and 2--10\,keV bands. In order to remove biases due to nonsimultaneity, only pairs of measurements that took place within 4\,hr are taken into account. Such a time window falls well below the minimum variability timescale measured in the VHE and X-ray regimes along the campaign. The results are shown in Fig.~\ref{VHE_vs_swift}. The blue data points highlight the measurements during the August 2019 flare, and the arrows show the corresponding direction of time. In order to quantify the correlation, we use the Pearson's coefficient as well as the discrete correlation function \citep[DCF; ][]{1988ApJ...333..646E}. The resulting values for each energy combination are listed in Table~\ref{tab:correlation_results_VHE_swift}.\par 

The results do not reveal any significant VHE versus X-ray correlation. In the case of $>300$\,GeV versus 0.3--2\,keV, the Pearson's coefficient is only $0.2_{-0.3}^{+0.2}$ with a low significance of $0.9\sigma$. Regarding $>300$\,GeV versus 2--10\,keV, the Pearson's coefficient and the significance are slightly higher ($0.6_{-0.2}^{+0.1}$, $2.7\sigma$), although the trend remains marginal. Along most of the campaign, the source is in a low state and the dynamic range of flux values is relatively small. This may be a main reason explaining the low level of correlation.\par 

The three consecutive measurements during the flare, highlighted with blue markers, shows an interesting correlated behaviour. Between the first and the third day of the flare, the VHE flux decayed by a factor of $\sim 3.5$. From, Fig.~\ref{VHE_vs_swift}, a hint of a simultaneous decay is also noticeable in the 2--10\,keV band. On the other hand, in the 0.3--2\,keV band the flux shows a roughly constant behaviour without any indication of a decay. Such different trends between two very nearby energy regions (0.3--2\,keV and 2--10\,keV) are difficult to reconcile with a single-zone model. In fact, it suggests that the VHE flare coincides with a region dominating the X-ray emission only at $\gtrsim2$\,keV, while the $\lesssim2$\,keV flux originates from a different emitting region possibly unrelated to the flare. In Sect.~\ref{flare_modelling}, we successfully interpret the flare using a two-component leptonic model in a time-dependent approach.

\begin{figure}[h!]
   \centering
   \includegraphics[width=1.0\columnwidth]{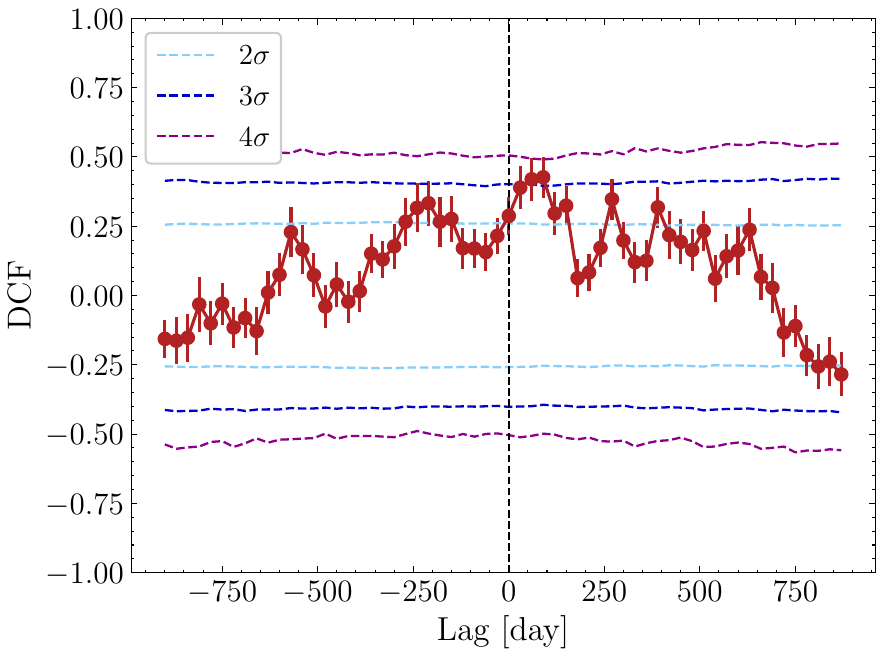}
   \caption{DCF between OVRO and \textit{Fermi}-LAT versus the time lag. Positive lags mean that the OVRO fluxes lag behind the \textit{Fermi}-LAT ones. The horizontal dashed lines represent the $2\sigma$ (light blue), $3\sigma$ (dark blue) and $4\sigma$ (magenta) confidence bands using Monte-Carlo simulations (see text).}
    \label{fermi_radio_dcf}
\end{figure}

\subsection{Radio versus MeV--GeV gamma-ray correlation}
\label{sec:radio_he_correlation_study}

For several blazars, studies have unveiled a positive correlation between the flux in the MeV--GeV and radio bands \citep[see, e.g.,][]{2008Natur.452..966M, 2010ApJ...722L...7P, 2014MNRAS.441.1899F}. The correlation typically occurs with a delay where the radio band lags behind the MeV--GeV with a time lag ranging from a few tens to a few hundreds of days \citep{2014MNRAS.445..428M}. We search for such correlation pattern in 1ES~2344+514 by taking advantage of the simultaneous long-term OVRO and \textit{Fermi}-LAT monitoring ($\sim 13$\,yr). The long-term light curves of OVRO and \textit{Fermi}-LAT are shown in Fig.~\ref{fermi_radio_lc} in Appendix~\ref{radio_fermi_longterm_lc}. The \textit{Fermi}-LAT fluxes are computed with monthly binning, in the same fashion as in Fig.~\ref{MWL_lightcurve}. We correlate the two bands and search for a time lag using the DCF \citep{1988ApJ...333..646E}. Regarding the \textit{Fermi}-LAT fluxes, only bins with a detection significance above $2\sigma$ (i.e., TS $>4$) are considered. The results are shown in Fig.~\ref{fermi_radio_dcf} in red markers. The DCF is evaluated in time-lag bins of 30\,days, being the smallest bin size of the two light curves. The highest DCF value ($\sim 0.43$) is found at a lag of $90$\,days. A positive lag means that the radio is delayed behind the MeV--GeV band.\par

We evaluate the significance of the DCF using dedicated Monte-Carlo simulations. First, the slope of the power spectral density \citep[PSD; see, e.g.,][]{2014MNRAS.445..437M} of the two light curves is estimated. For this we adopted the same approach as the one described by \citet{2021A&A...655A..89M}. We obtain a PSD power-law index of $\beta_{LAT}=-0.9$ for the \textit{Fermi}-LAT data, and $\beta_{\rm OVRO}=-2.1$. Those values match the one generally found in BL Lac type objects \citep{2014MNRAS.445..428M}. Using these PSD shapes, $10^5$ \textit{realistic} and uncorrelated light curves are simulated with the same temporal sampling and binning as the data. From this set of simulations, the distribution of the DCF is extracted in each lag bin to derive $2\sigma$, $3\sigma$, and $4\sigma$ confidence bands. They are plotted in cyan, blue, and magenta dashed lines, respectively. We note that the light curves are simulated using the \citet{1995A&A...300..707T} method, which is valid only for Gaussian-distributed fluxes. This assumption is valid for the OVRO data. For the \textit{Fermi}-LAT light curve, the fluxes are not exactly Gaussian-distributed. Because of that, we also simulated the light curves using the prescription of \citet{2013MNRAS.433..907E}, which preserves any underlying flux distribution. However, the results do not show a significant difference. In conclusion, our simulations indicate that the correlation is at the level of $3\sigma$.\par 

In order to estimate the time-lag uncertainty, we follow the common method of \citet{1998PASP..110..660P} and \citet{2004ApJ...613..682P}. We generate 2000 pairs of light curves in both energy bands using flux randomisation and random subset selection. For each of these light-curve pairs, the DCF is computed and the centroid of the lags above 80\% of the maximum DCF value is calculated. The centroid is computed within a range of 320\,days around the peak seen at $\tau_{\rm lag}\approx90$\,days in Fig.~\ref{fermi_radio_dcf}. Finally, the $1\sigma$ confidence band for the lag is estimated from the 68\% containment of the centroid distribution. Using this method, we obtain a $1\sigma$ confidence band of $\tau_{\rm lag}\in [30, 106]$\,days.  

\begin{figure*}[h!]
   \centering
   \includegraphics[width=2.0\columnwidth]{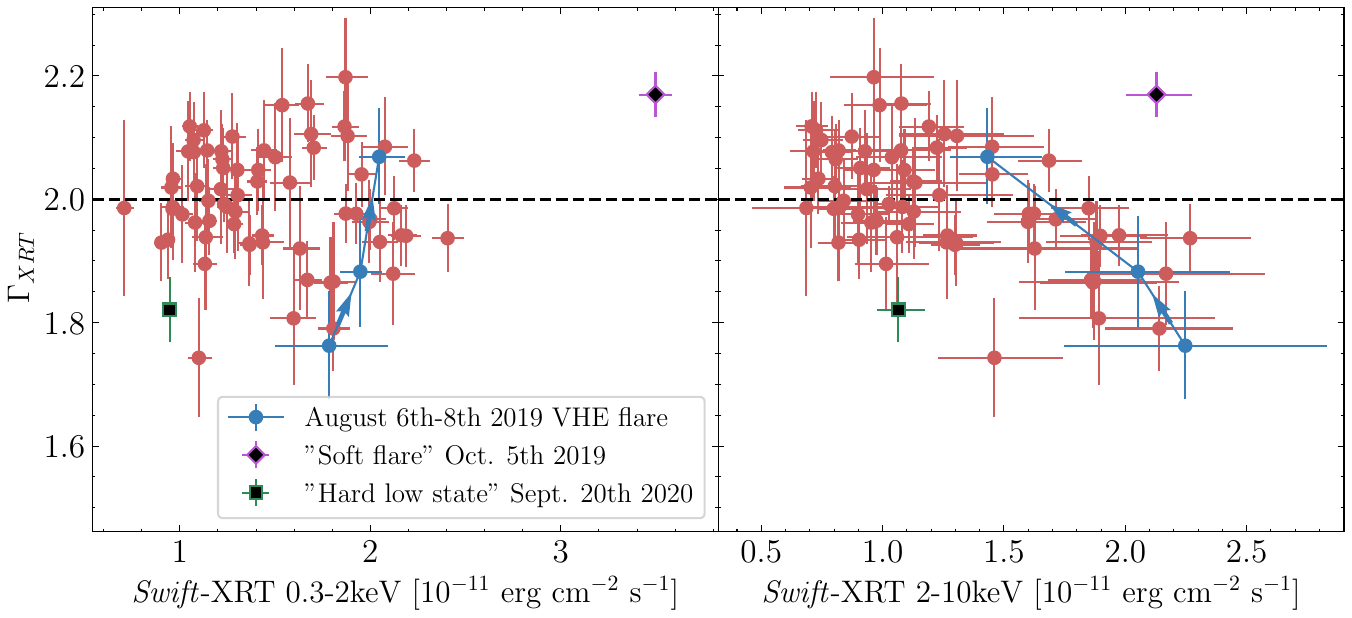}
   \caption{X-ray power-law index versus flux over the multiwavelength campaign using \textit{Swift}-XRT data. Fluxes are nightly binned. The blue measurements correspond to the flaring state in August 2019 and the arrows show the direction of time. The green and violet markers correspond to the ``soft flare'' (October 5\textsuperscript{th} 2019) and ``hard low state'' (September 20\textsuperscript{th} 2020) periods.}
    \label{index_versus_flux_swift}
\end{figure*}

\begin{figure}[h!]
   \hspace{-0.85cm}
   \includegraphics[width=1.1\columnwidth]{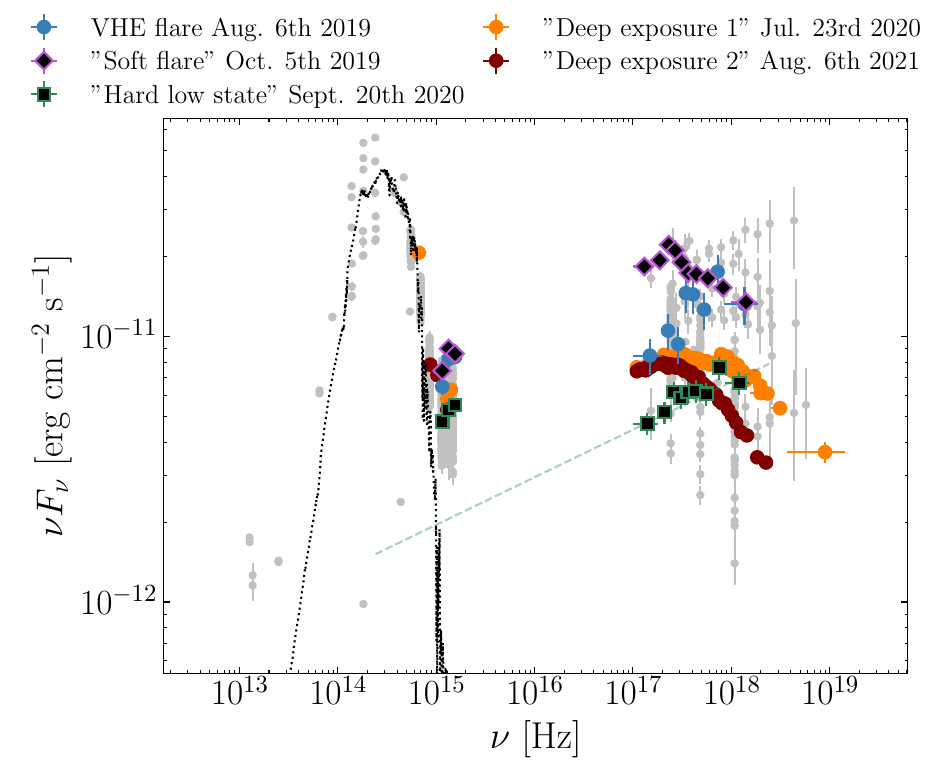}
   \caption{\small Optical/UV and X-ray emission during specific states. The violet markers depict the \textit{Swift} spectrum during a soft flare on October 5\textsuperscript{th} 2019, during which the power-law index is above 2, indicating a synchrotron peak energy below 1\,keV. The green markers display the \textit{Swift} spectrum during a low but hard state on September 20\textsuperscript{th} 2020. During the latter day, the source is in an EHBL state despite the particularly low flux. An extrapolation of the X-ray best-fit power-law model to lower energy (shown with a green dashed line) suggests a ``UV excess'' and possibly indicates a second component contributing to the synchrotron SED. The orange and maroon markers correspond to the ``deep exposure 1'' and ``deep exposure 2'' SEDs. The blue markers represent the \textit{Swift} spectrum during the VHE flare on August 6\textsuperscript{th} 2019. Grey points are archival measurements from the SSDC database for comparison purposes. The black dotted line is the host-galaxy template, also taken from the SWIRE database.}
    \label{synchrotron_remarkable_day}
\end{figure}

\section{Spectral evolution and study of the intermittent EHBL behaviour}
\label{sec:spectral_study}

\subsection{X-ray spectral evolution} 
\label{sec:X-ray_spectral_study}



1ES~2344+514 is known for its intermittent EHBL behaviour in the synchrotron domain, with a synchrotron peak frequency shifting above 1\,keV (i.e., $\gtrsim2.4\times 10^{17}$\,Hz) temporarily. The shift seems to occur primarily during flaring episodes, as observed in 1996 by \citet{2000MNRAS.317..743G} and more recently in 2016 by \citet{2020MNRAS.496.3912M}. This behaviour follows the common ``harder-when-brighter'' trend observed in other BL Lac type objects \citep{2004ApJ...601..151K, 2021MNRAS.504.1427A, 2021A&A...655A..89M}. With the \textit{Swift}-XRT observations gathered during the multiwavelength campaign presented in this work we study the X-ray spectral variability over a $\sim 3$\,yr time frame and characterise in deeper detail the occurrence of intermittent EHBL states.\par

In Fig.~\ref{index_versus_flux_swift} we show the \textit{Swift}-XRT power-law index $\Gamma_{\rm XRT}$ versus the 0.3--2\,keV and 2--10\,keV fluxes over the entire campaign. $\Gamma_{\rm XRT}$ is obtained by fitting the power-law model $dN/dE \propto E^{-\Gamma_{\rm XRT}}$ over the 0.3--10\,keV range. We note that 68 out of 72 \textit{Swift}-XRT fits have a $\chi^2$ yielding a $p$-value below $2\sigma$, indicating that a simple power-law model provides a satisfactory description of the spectrum in the vast majority of cases. Only two observations reveal a preference for a log-parabolic model at a significance above $3\sigma$. For the 0.3--2\,keV band, the spectral hardness does not show any hint of correlation with the flux. On the other hand, in the 2--10\,keV band evidence of a ``harder-when-brighter'' trend is found. The Pearson's correlation coefficient is $-0.5 \pm 0.1$ with a significance of $4.6\sigma$. The three days during the 2019 flare at VHE are highlighted in blue, and show a softening of the emission throughout the flux-decaying phase. The fact that the ``harder-when-brighter'' trend is mostly visible when looking at the 2--10\,keV band may be due to a larger variability in the latter band compared to the 0.3--2\,keV band. For completeness, we list in Appendix~\ref{sec:swift_xrt_parameters} the spectral results of each \textit{Swift}-XRT observation both using a power-law and a log-parabola model. \par

A significant fraction of the measurements have $\Gamma_{\rm XRT}<2$, suggesting a synchrotron peak frequency located above or around 1\,keV, in agreement with an EHBL behaviour. The hint of anticorrelation in the right panel of Fig.~\ref{index_versus_flux_swift} suggests that an increased 2--10\,keV flux generally correlates with an EHBL state. Nevertheless, throughout the campaign several outliers exist pointing toward a more complex phenomenology. In Fig.~\ref{index_versus_flux_swift}, we highlight two emblematic outliers. The first one is a soft X-ray flare that occurred on October 5\textsuperscript{th} 2019 (MJD~58761) and it is plotted with a magenta diamond marker. The second one is a hard low state on September 20\textsuperscript{th} 2020 (MJD~59112), plotted with a green square marker.\par

The ``soft X-ray flare'' is identified by a 0.3--2\,keV flux of $\sim 3.5 \times 10^{-11}$\,erg\,cm$^{-2}$\,s$^{-1}$, which is by far the highest flux measured in this band over the campaign. Also, the 2--10\,keV state is among the highest measured. On the other hand, it shows one of the softest states during the campaign and $\Gamma_{\rm XRT} \approx 2.17$, implying a synchrotron peak below 1\,keV despite the high state. The measurement clearly stands out from the rest of the points in Fig~\ref{index_versus_flux_swift}. It is worth mentioning that on that day, a simultaneous MAGIC measurement yields a flux above 300\,GeV around 8\% of the Crab Nebula, which is twice the nonflaring state of 1ES~2344+514 \citep{2017MNRAS.471.2117A}. The analysis of \textit{Fermi}-LAT data averaged over one day centered around October 5\textsuperscript{th} 2019 yields a $\sim 4.3\sigma$ detection and a flux in the 0.3--300\,GeV band of $(5.2\pm2.9)\times10^{-8}$\,cm$^{-2}$s$^{-1}$. This is about 5 times the monthly average around that date, although the relatively large uncertainty prevents us from firmly claiming the presence of an MeV--GeV flare. In any case, such a flare would not be surprising since in leptonic models the electrons emitting in the 0.3--2\,keV band are also responsible for the MeV--GeV flux \citep{1998ApJ...509..608T}. \par 


Regarding the ``hard low state'' day, both the 0.3--2\,keV and 2--10\,keV fluxes are relatively low, around $10^{-11}$\,erg\,cm$^{-2}$\,s$^{-1}$. On the other hand, it shows one of the hardest spectra with $\Gamma_{\rm XRT} \approx 1.8$. The source is thus in an EHBL state despite the low activity. In fact, the X-ray spectrum exhibited a similar hardness as during the VHE flare. A VHE measurement performed a few hours after \textit{Swift} reveals a low flux of 4\% that of the Crab Nebula. The X-ray spectral evolution during those two peculiar epochs goes against the typical ``harder-when-brighter'' relation. An EHBL state in 1ES~2344+514 thus occurs independently from the flux activity. We note that a comparable ``hard low state'' is visible in Fig.~\ref{synchrotron_remarkable_day} at $\Gamma_{\rm XRT}\approx1.75$ with a 0.3--2.0\,keV flux of $\sim 10^{-11}$\,erg\,cm$^{-2}$\,s$^{-1}$. The later measurement, which has one of the lowest 0.3--2\,keV fluxes, indicates that an EHBL state in low activity is a feature that repeats over time. No simultaneous MAGIC data are available.\par

In Fig.~\ref{synchrotron_remarkable_day}, the \textit{Swift}-XRT SEDs for those two selected outliers are plotted, together with the \textit{Swift}-UVOT measurements to obtain a comprehensive view of the synchrotron component. The flare state on August 6\textsuperscript{th} 2019 is also plotted with blue markers. Fig.~\ref{synchrotron_remarkable_day} emphasizes a strong X-ray spectral variability, inconsistent with the ``harder-when-brighter'' evolution (as already mentioned).\par

\subsubsection{Evidence of a ``UV'' excess}
Another interesting result from Fig.~\ref{synchrotron_remarkable_day} stems from an apparent mismatch between the X-ray and UV fluxes during the ``hard low state'' (green markers). In other words, the extrapolation to lower energies of the \textit{Swift}-XRT spectrum falls significantly below the \textit{Swift}-UVOT data points (by a factor $\sim 2.5$). In Fig.~\ref{synchrotron_remarkable_day}, a green dashed line shows the extrapolation to lower energies of the best-fit power-law model. The corresponding X-ray power-law index is significantly below 2 ($\Gamma_{\rm XRT}=1.82\pm0.05$), while the energy flux around the low-energy \textit{Swift}-XRT points is at the level or even below the \textit{Swift}-UVOT measurements. In addition to that, the fluxes in the three \textit{Swift}-UVOT filters point toward a relatively hard spectrum in the UV, further confirming the mismatch between the \textit{Swift}-XRT and \textit{Swift}-UVOT data. Using a two-point photon index formula, $\Gamma_{\rm UV}=-\frac{\log{ \left(F_{UVW1}/F_{UVW2}\right) }}{\log\left(\nu_{UVW1}/\nu_{UVW2}\right)} + 1$ \citep[see e.g.][]{2015A&A...575A..13F}, we estimate the spectral index in the UV to be $\Gamma_{\rm UV}=1.7\pm0.4$ (we neglect in this computation the host-galaxy contribution). For the other days shown in Fig.~\ref{synchrotron_remarkable_day}, the evidence for an excess is less apparent. The combination of a hard and particularly low X-ray state on September 20\textsuperscript{th} 2020 renders its detection easier. The other ``hard low state'' visible in Fig.~\ref{synchrotron_remarkable_day} mentioned in the previous section, which is characterised by $\Gamma_{\rm XRT}\approx1.75$ and a 0.3--2.0\,keV flux of $\sim 10^{-11}$\,erg\,cm$^{-2}$\,s$^{-1}$, displays the same mismatch between the UV data and the \textit{Swift}-XRT spectrum. \par 

The contribution from the host galaxy plotted as a black dotted line in Fig.~\ref{synchrotron_remarkable_day} is negligible in the UV band \citep[$\lesssim10\%$;][]{2014MNRAS.442..629R}, in particular for the UVOT $W2$ filter, and cannot explain the ``UV excess''. We consider here a host template of a 13\,Gyr old elliptical galaxy taken from the SWIRE\footnote{\url{http://www.iasf-milano.inaf.it/~polletta/templates/swire_templates.html}} library \citet{2007ApJ...663...81P}. Using different elliptical templates, like the one from \citet{1996ApJ...467...38K}, indicates a similar host contribution in the UV band, and thus leads to the same conclusions. Overall, such a spectral feature indicates two different electron populations contributing to the UV and X-ray emission of 1ES~2344+514 (see Sect.~\ref{low_state_model} and Sect.~\ref{sec:discussion}).\par

\subsubsection{\textit{XMM-Newton}, \textit{NuSTAR}, and \textit{AstroSat} deep exposures}
\label{sec:X_ray_spectral_deep_exposures}

This section presents the spectral results of the long X-ray exposures that took place simultaneously with multi-hour MAGIC observations, ``deep exposure 1'' and ``deep exposure 2''. The first one took place on July 23\textsuperscript{rd} 2020 and includes X-ray data from \textit{XMM-Newton} and \textit{NuSTAR}. The second exposure includes \textit{XMM-Newton} and \textit{AstroSat-SXT} X-ray data and happened on August 6\textsuperscript{th} 2021. Table~\ref{tab:XMM_nustar_exposures} summarizes the exposures for each instrument. The combination of \textit{XMM-Newton}, \textit{NuSTAR}, and \textit{AstroSat} provides an unprecedented characterisation of the synchrotron emission around the SED peak frequency for 1ES~2344+514.\par

\begin{table*}[h!]
\caption{\label{tab:XMM_nustar_spectral_param} X-ray spectral parameters from \textit{XMM-Newton}, \textit{NuSTAR}, and \textit{AstroSat-SXT} observations using a log-parabola model during the deep exposures with MAGIC.} 
\renewcommand{\arraystretch}{1.1}
\centering
\begin{tabular}{ l c c c c c c}     
\hline\hline 
 Epoch/Date & Instruments  & flux   & $\alpha$ & $\beta$ & $\chi^{2}$/dof & $\nu_{\rm synch,p}$\\  
 &  & [$10^{-11}$\,erg\,cm$^{-2}$\,s$^{-1}$]  &  &  & & [keV]\\  
\hline\hline   
 Deep exposure 1  &  &  &  &  &  \\
 July 23\textsuperscript{rd} 2020 & \textit{XMM-Newton}   & $1.06\pm0.01$ & $1.94 \pm 0.01$ & $0.26 \pm 0.02$ & 476.7/449 & $1.32 \pm 0.04$ \\
& \textit{XMM-Newton} $+$ \textit{NuSTAR}   & $1.05\pm0.01$ & $1.94 \pm 0.01$ & $0.25 \pm 0.02$ & 794.6/782 & $1.33 \pm 0.04$ \\
\hdashline

Deep exposure 2 &  &  &  &  & & \\
August 6\textsuperscript{th} 2021  & \textit{XMM-Newton}    & $0.75\pm0.01$ & $2.07\pm0.01$ & $0.41\pm0.02$ & 518.4/429 & $0.82 \pm 0.03$\\
& \textit{AstroSat-SXT}    & $0.93\pm0.02$ & $2.01\pm0.14$ & $0.51\pm0.27$ & 108/113 & $0.97 \pm 0.12$\\

\hline 

\end{tabular}
\tablefoot{In the case of \textit{XMM-Newton} and \textit{XMM-Newton}+\textit{NuSTAR} fits, the fluxes are given in the 2--10\,keV band. The \textit{AstroSat-SXT} flux is computed in the 0.7--7\,keV band. The normalisation energy is 1\,keV and the Galactic column density of hydrogen is fixed to $N_{\rm H} = 1.41\times10^{21}$\,cm$^{-2}$.} 
\end{table*}

\begin{table*}[h!]
\caption{\label{tab:MAGIC_spectral_param} MAGIC spectral parameters for epochs of interest obtained from a power-law fit ($dN/dE = f_0 (E/E_0)^{-\Gamma_{\rm VHE}}$) between 100\,GeV and 2\,TeV.} 
\renewcommand{\arraystretch}{1.15}
\centering
\begin{tabular}{ l c c c c c c c }     
\hline\hline 
 Epoch & Date &  & Flux $>300$\,GeV & $f_{0}$ & $\Gamma_{\rm VHE}$ & $\chi^{2}$/dof \\  
 &  &  & $[10^{-11} \mathrm{cm}^{-2} \mathrm{s}^{-1}]$ & $[10^{-11} \mathrm{cm}^{-2} \mathrm{s}^{-1} \mathrm{TeV}^{-1}]$ &  & \\  
\hline\hline   
& August 6\textsuperscript{th} 2019 &  & $2.58\pm0.47$ & $3.5\pm0.6$ & $1.9 \pm 0.2$ & 7.3/3 \\
VHE flare period & August 7\textsuperscript{th} 2019 &  & $1.55\pm0.38$ & $2.3\pm0.5$ & $2.5 \pm 0.2$ & 7.3/4 \\ 
& August 8\textsuperscript{th} 2019 &  & $0.73\pm0.30$ & $1.0\pm0.4$ & $2.5 \pm 0.5$ & 0.1/1 \\

\hdashline

``Soft X-ray flare'' & October 5\textsuperscript{th} 2019 &  & $0.98\pm0.31$ & $1.3\pm0.4$ & $1.8 \pm 0.3$ & 3.4/3 \\
``Hard low X-ray state'' & September 21\textsuperscript{st} 2020 &  & $0.54\pm0.23$ & $0.5\pm0.3$ & $2.1 \pm 0.5$ & 3.2/1 \\

\hline
Deep exposure 1 & July 23\textsuperscript{rd} 2020 &  & $0.46\pm0.16$ & $0.7\pm0.2$ & $2.5 \pm 0.3$ & 7.8/3 \\
Deep exposure 2 & August 6\textsuperscript{th} 2021 &  & $0.26\pm0.12$ & $0.3\pm0.2$ & $2.4 \pm 0.4$ & 1.8/1 \\
\hline 
\hline
2020 Average &  &  & $0.44\pm0.06$ & $0.7\pm0.1$ & $2.4 \pm 0.1$ & 6.7/4 \\
\hline
\hline
2021 Average &  &  & $0.21\pm0.07$ & $0.3\pm0.1$ & $2.3 \pm 0.4$ & 1.5/3 \\
\hline
\end{tabular}
\tablefoot{\centering The normalisation energy is 500\,GeV and the data are corrected for EBL absorption using the EBL template of \citet{2011MNRAS.410.2556D}.} 
\end{table*}

The spectra are fitted with a fixed Galactic column density of hydrogen ($N_{\rm H} = 1.41\times10^{21}$\,cm$^{-2}$) using a power-law and a log-parabola model:  $dN/dE \propto  (E/E_0)^{-\Gamma}$ and $dN/dE \propto  (E/E_0)^{-\alpha-\beta \log{E/E_0}}$, with $E_0=1$\,keV as the pivot energy. For all instruments, we average the spectrum over the entire exposure time given the low flux variability and the absence of significant intraday spectral variability. The \textit{XMM-Newton} spectral parameters are derived by simultaneously fitting the EPIC PN, MOS1, and MOS2 data. Regarding the ``deep exposure 1'' epoch, the parameters are obtained from \textit{XMM-Newton}-only data as well as the combined \textit{XMM-Newton}+\textit{NuSTAR} spectra given that the two instruments complement each other in energy. In all cases and for both nights, from the \textit{XMM-Newton} fits the power-law model is significantly rejected ($>5\sigma$), implying the detection of a curvature in the 0.3--10\,keV range. In what follows, we thus only consider and discuss the log-parabola model. \par 

The results of the fits are listed in Table~\ref{tab:XMM_nustar_spectral_param}. The last column is the extracted synchrotron peak frequency, $\nu_{\rm synch,p}$, that is obtained in \texttt{XSPEC} using the \texttt{eplogpar} model (which is essentially the same function as the log parabola defined earlier, but refactored such that $\nu_{\rm synch,p}$ becomes an explicit parameter of the model). ``Deep exposure 1'', where the 2--10\,keV flux is slightly higher, shows a harder $\alpha$ with respect to the ``deep exposure 2'' epoch ($\alpha=1.94\pm0.01$ versus $\alpha=2.07\pm0.01$ from the \textit{XMM-Newton} data). In addition, the curvature $\beta$ is significantly more pronounced during ``deep exposure 2''. Such ``harder-when-brighter'' evolution is confirmed by the evolution of $\nu_{\rm synch,p}$. The fits yield a small, but significant shift of $\nu_{\rm synch,p}$ between the two epochs. We find $\nu_{\rm synch,p}=1.32 \pm 0.04$\,keV for ``deep exposure 1'' while during ``deep exposure 2''  \textit{XMM-Newton} and  \textit{AstroSat} consistently unveil a lower value: $\nu_{\rm synch,p}=0.82 \pm 0.03$\,keV for \textit{XMM-Newton} and $\nu_{\rm synch,p}=0.97 \pm 0.12$\,keV using \textit{AstroSat}. It is worth noting that despite the low activity $\nu_{\rm synch,p}$ is larger than 1\,keV for ``deep exposure 1'', thus within the EHBL family according to the definition of \citet{2001A&A...371..512C}. In conclusion, similarly to what is discussed in the previous section, 1ES~2344+514 can show EHBL behaviour also in low states. In Fig.~\ref{synchrotron_remarkable_day} we show the \textit{XMM-Newon} and \textit{NuSTAR} SEDs. For \textit{XMM-Newton} only the data from EPIC-pn camera are used to build the SED for simplicity and also given the larger number of counts compared to the ones in the MOS1 and MOS2 cameras. The UV data at $\nu>10^{15}$\,Hz  (obtained from the \textit{XMM-Newton} OM instrument) receive negligible contribution from the host galaxy \citep{2014MNRAS.442..629R}. \par

Regarding ``deep exposure 1'' the difference in the parameters between the \textit{XMM-Newton}-only and \textit{XMM-Newton}/\textit{NuSTAR}-combined fits is not significant. This indicates a smooth connection between the soft X-ray band (up to $\sim10$\,keV) covered by \textit{XMM-Newton} and the hard X-ray band covered by \textit{NuSTAR} ($\gtrsim 10$\,keV). We also stress that the cross-calibration factors (derived from the fits in \texttt{Xspec}) between \textit{XMM-Newton} and \textit{NuSTAR} are all below 15\%, thus within the systematics of the scientific instrumentation onboard these two spacecrafts \citep{2017AJ....153....2M}. For the ``deep exposure 2'' epoch, the \textit{AstroSat-SXT} spectral parameters (derived in the 0.7--7\,keV range) are consistent with the ones obtained in the 0.4--10\,keV band by \textit{XMM-Newton}.

\subsection{VHE spectral evolution}
\label{sec:magic_spectral_study}

The MAGIC observations are used to probe the spectral variability in the VHE band. Unfortunately, given the low flux, the VHE spectral slope cannot be resolved with high resolution on a single snapshot for most of the MAGIC observing nights. A meaningful VHE spectral study on a daily timescale over the full campaign is thus not possible (unlike in the X-ray with \textit{Swift}-XRT, \textit{XMM-Newton}, \textit{NuSTAR}, and \textit{AstroSat}). Thus, we limit the VHE spectral study to a few specific epochs of interest: the VHE flaring period, the ``soft X-ray flare'', and the deep exposures nights simultaneous with \textit{XMM-Newton}, \textit{NuSTAR}, and \textit{AstroSat}. We also compute a VHE spectrum using a MAGIC observation close to the ``hard low state'' in the X-ray discussed earlier. The corresponding MAGIC observation is not strictly simultaneous with the \textit{Swift}-XRT one, but took place roughly 14\,hr after. All spectra are fitted between 100\,GeV and 2\,TeV with a simple power-law spectrum, $dN/dE = f_0 (E/E_0)^{-\Gamma_{\rm VHE}}$, with $E_0=500$\,GeV the normalisation energy. This simple function provides a satisfactory description of the data, and the preference for a log-parabola model (one additional degree of freedom) is always below $3\sigma$. All parameters are computed after correcting the spectra for extragalactic background light (EBL) absorption using the template EBL model from \citet{2011MNRAS.410.2556D}.\par  

Table~\ref{tab:MAGIC_spectral_param} lists the resulting best-fit parameters. Evidence of daily timescale spectral variability is measured between the first and the second days of the flare. On August 6\textsuperscript{th} 2019, the brightest day of the flare, the spectrum is hard with $\Gamma_{\rm VHE} = 1.9 \pm 0.2$. On the night after, August 7\textsuperscript{th} 2019, the spectrum steepens to $\Gamma_{\rm VHE} = 2.5 \pm 0.2$ together with the flux. This ``harder-when-brighter'' trend closely follows what is also measured in the X-ray (see blue points in Fig.~\ref{index_versus_flux_swift} and Sect.~\ref{sec:X-ray_spectral_study}). During the flaring event in 2016 discussed by \cite{2020MNRAS.496.3912M}, the power-law index is around 2.0--2.1 (depending on the exact day), being consistent with what we report here for the peak activity of the August 2019 flare. Regarding the ``soft X-ray flare'' night, the spectrum is similarly hard ($\Gamma_{\rm VHE} = 1.8 \pm 0.3$), although the VHE flux is close to the quiescent state (8\% that of the Crab Nebula). Regarding the night close to the ``hard low state'' in the X-ray band, the uncertainty of the slope ($\Gamma_{\rm VHE} = 2.1 \pm 0.5$) does not allow us to make any strong claim about the VHE spectral shape. \par

During the deep exposure nights (together with \textit{XMM-Newton}, \textit{NuSTAR}, and \textit{AstroSat}), $\Gamma_{\rm VHE}$ is around 2.4--2.5 and remains consistent within statistical uncertainties. Differently from the X-ray (see previous section), no VHE spectral variability is measured during those nights, which may also be due to the limited sensitivity of MAGIC compared to \textit{XMM-Newton}, \textit{NuSTAR}, and \textit{AstroSat}.\par 

In Table~\ref{tab:MAGIC_spectral_param} we also show the power-law index derived for the entire 2020 year, during which we do not find any significant flare. The best-fit index is $\Gamma_{\rm VHE} = 2.4 \pm 0.1$, similar to the deep exposure nights as well as to the days following the peak VHE activity on August 6\textsuperscript{th} 2019. Within (statistical and systematic) uncertainties, such an average slope is consistent with previous measurements during quiescent states \citep{2007ApJ...662..892A, 2013A&A...556A..67A, 2017MNRAS.471.2117A}. During 2021, the total detection significance is only $\sim 2\sigma$, preventing a precise determination of the average spectral hardness. The corresponding best-fit index is $\Gamma_{\rm VHE} = 2.3 \pm 0.4$, consistent with the 2020 average. \par 

Overall, the MAGIC observations unveil VHE spectral variability, although moderate. Except during the August 2019 VHE flare and ``soft X-ray flare'' in October 2019, the measured power law is typically around $\Gamma_{\rm VHE} \approx 2.4$--2.5 (consistent with published work during quiescent states).\par 

In Appendix~\ref{sec:magic_spectral_fit}, we overlay the best-fit MAGIC power-law models in order to better appreciate the spectral variability.

\section{Multiwavelength characterization of the quiescent activity and its broad-band modelling}
\label{low_state_model}

\begin{figure*}[h!]
    \centering
    \begin{subfigure}[b]{0.497\textwidth}
        \centering
        \includegraphics[width=\textwidth]{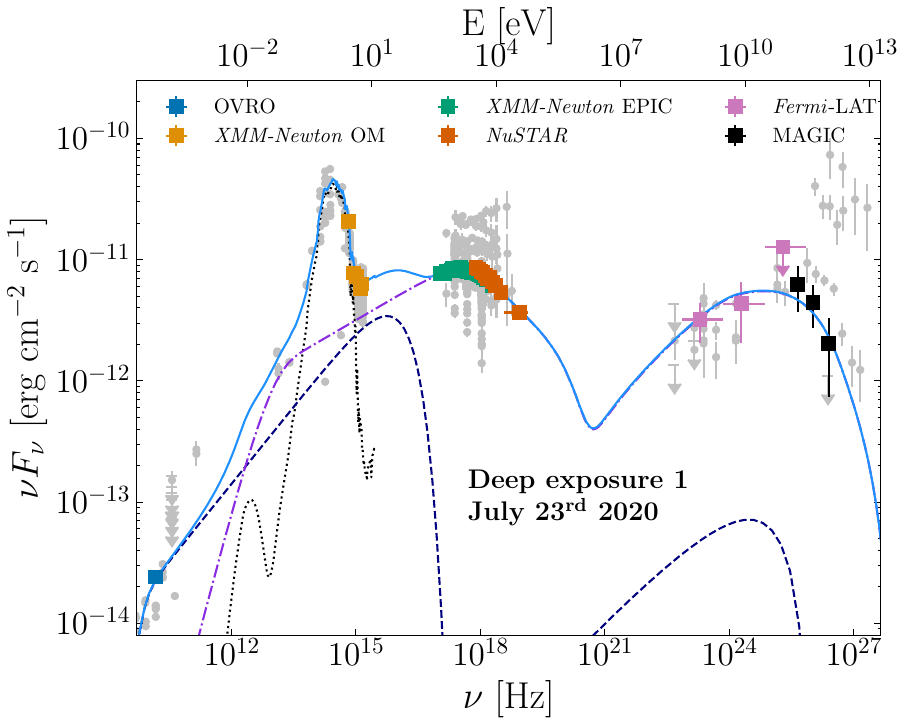}
    \end{subfigure}
    \begin{subfigure}[b]{0.497\textwidth}  
        \centering 
        \includegraphics[width=\textwidth]{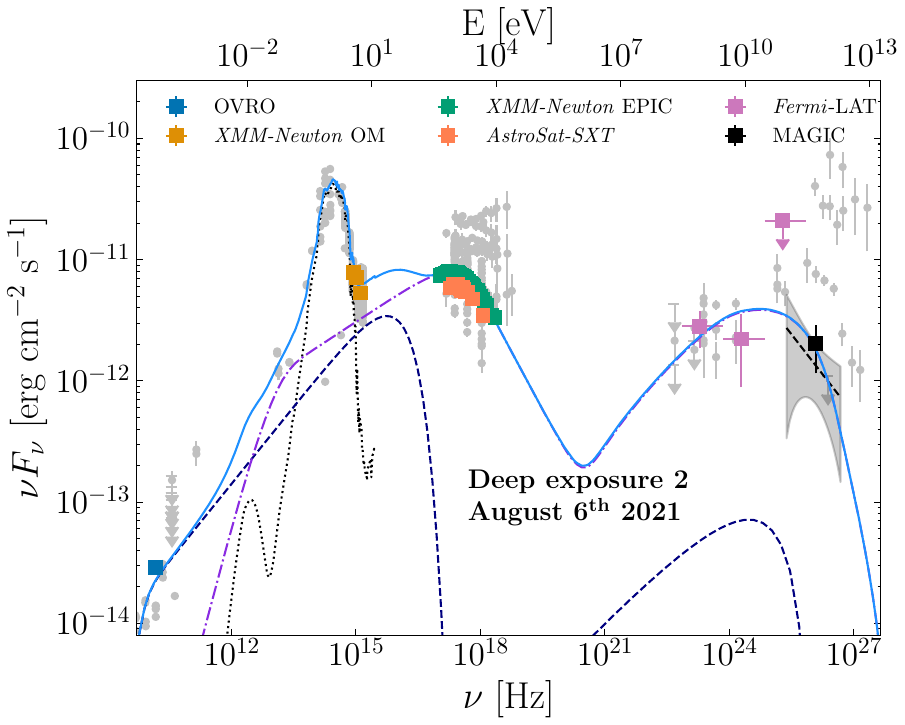}
    \end{subfigure}
    \caption{Broad-band SEDs during the simultaneous MAGIC, \textit{XMM-Newton}, \textit{NuSTAR}, and \textit{AstroSat} observations. The panel on the left shows the SED from the ``deep exposure 1'' epoch (July 23\textsuperscript{rd} 2020) while the panel on the right is during the ``deep exposure 2'' epoch (August 6\textsuperscript{th} 2021). The results of the 2-component SSC modelling is also shown: the dashed dark-blue line is the emission from the ``core'' region, while the violet dash-dotted line is the emission from the ``variable'' region (see text for more details). The sum of the two components is plotted in a continuous light-blue line. The parameter values of the model are listed in Table~\ref{tab:ssc_parameters_deep_exposures}. The large bump in the optical domain (black dashed curve) is the emission from the host galaxy modelled with the template from the SWIRE database \citep{2007ApJ...663...81P}. As in Fig.\ref{synchrotron_remarkable_day}, grey points are archival measurements from the SSDC database.
} 
    \label{deep_exposures_nights_sed}
\end{figure*}

Fig.~\ref{deep_exposures_nights_sed} shows the broad-band SEDs during the ``deep exposure 1'' and ``deep exposure 2'' epochs. The corresponding VHE emission is at the level of $\sim$4\% and $\sim$2\% that of the Crab Nebula above 300\,GeV, respectively (see Table~\ref{tab:MAGIC_spectral_param}). The 2-10\,keV fluxes are close to $10^{-11}$\,erg\,cm$^{-2}$\,s$^{-1}$ (see Table~\ref{tab:XMM_nustar_spectral_param}). As highlighted in the Sect.~\ref{introduction_section}, such flux levels correspond to the quiescent activity of the source. The MAGIC spectra are corrected for the EBL absorption effects. In view of the low detection significance by MAGIC for ``deep exposure 2'', the corresponding VHE SED contains only one point with a $\sim 2\sigma$ signal. We thus complement the VHE spectrum with a ``butterfly'' envelope indicating the $1\sigma$ statistical uncertainty on the spectral shape derived from the power-law fit. In the radio, we use the closest OVRO measurements, which took place $\lesssim 1$\,day away from the one of MAGIC but can be assumed as simultaneous given the low variability at such energies (see Sect.~\ref{MWL_variability}). For the UV data, we consider the fluxes from the \textit{XMM-Newton} OM instrument that were obtained simultaneously with the X-ray measurements. Regarding \textit{Fermi}-LAT, the SEDs are averaged over 1 month around the ``deep exposure 1'' observation and over 2 months around ``deep exposure 2''. Such integration times are needed to achieve a significant detection (TS $> 25$). The grey data points are the archival measurements retrieved from the SSDC. A comparison with the archival data reveals that the source is probed in one of its lowest VHE gamma-ray states measured so far. For ``deep exposure 2'', the X-ray flux is also among the lowest.\par

\begin{table*}[h!]
\caption{\label{tab:ssc_parameters_deep_exposures}Parameters of the two-components SSC models obtained for the MAGIC, \textit{XMM-Newton}, \textit{NuSTAR}, and \textit{AstroSat} simultaneous observing nights.}
\centering
\begin{tabular}{l c c c @{\hskip 0.3in} c}     
\hline\hline
   & \multicolumn{2}{c}{Variable region} &  & Core region \\\cline{2-3} 
Parameters   & ``deep exposure 1'' & ``deep exposure 2'' & & \\
   & July 23\textsuperscript{rd} 2020 & August 6\textsuperscript{th} 2021 & & \\
\hline
\hline
$B'$ [$10^{-2}$\,G]  & 5.5 & 6.3 & & 5.0  \\
$R'$ [$10^{16}$\,cm] & 2 & 2 & & 10 \\
$\delta$ & 10 & 10 & & 10\\
$U'_{\rm e}$ [$10^{-3}$\,erg cm$^{-3}$] & 6.9 & 4.9 & & 0.1\\
$n_1$ & 2.60 & 2.55 & & 2.15\\
$n_2$ & 3.66 & 4.29 & & --\\
$\gamma'_{\rm min}$ & $2\times10^{3}$ & $2\times10^{3}$ & & 10\\
$\gamma'_{\rm br}$ & $4.9\times10^{5}$ & $3.7\times10^{5}$ & & --\\
$\gamma'_{\rm max}$ & $8\times10^{6}$ & $8\times10^{6}$ & & $9.3\times10^{4}$\\

\hline
\end{tabular}
\tablefoot{See text in Sect.~\ref{low_state_model} for the description of each parameter.} 
\end{table*}

The extensive multiwavelength coverage allows us to model the emission assuming a leptonic scenario \citep{1992ApJ...397L...5M, 1998ApJ...509..608T, 2004ApJ...601..151K} in order to constrain the physical properties of the quiescent state of 1ES~2344+514. The leptonic scenario considered here involves synchrotron radiation by a population of relativistic electrons as well as IC scattering off the synchrotron photons (the so-called SSC model). The particle interaction processes are computed using routines from the JetSeT software \citep{2006A&A...448..861M, 2009A&A...501..879T, 2011ApJ...739...66T, 2020ascl.soft09001T}.\par

In light of the results presented in Sect.~\ref{sec:X-ray_spectral_study}, which provides strong evidence that two separate particle populations contribute to the synchrotron emission, we consider a two-component model. The two components consist of two spherical regions homogeneously filled with electrons that are spatially separated and, thus, not interacting with each other. Further, we assume that each component is embedded in a homogeneous magnetic field. One region, dubbed as ``variable'', dominates the emission from optical to VHE. The second region, that we call ``core'', dominates mostly in the radio but brings a quite relevant contribution to the IR/optical/UV spectrum such that it is responsible for the ``UV excess'' reported in Sect.~\ref{sec:X-ray_spectral_study}. For simplicity, we assume that the two components are streaming down the jet with the same speed, and both having a Doppler factor of $\delta=10$. This value is in reasonable agreement with those derived by \citet{2009A&A...494..527H} and \citet{2021ApJ...923...30L} with Very Long Baseline Interferometry (VLBI) observations, from the variability
brightness temperature in the radio \citep{2018ApJ...866..137L},  as well as those obtained from SED modelling of BL Lac type objects \citep{2010MNRAS.401.1570T}.\par 

The electron distribution in the ``variable'' component is modelled as a broken power-law distribution,

\begin{equation}
    \frac{dN'}{d\gamma'}(\gamma')= \begin{cases}
    N'_0\, \gamma'{}^{-n_1}, \quad \gamma'_{\rm min}<\gamma'<\gamma'_{\rm br}\\
    N'_0\, \gamma_{\rm br}'{}^{n_2-n_1} \gamma'^{-n_2}, \quad \gamma'_{\rm br}<\gamma'<\gamma'_{\rm max}\, ,\\
    \end{cases}
\end{equation}
\noindent
where $N'_0$ is a normalisation constant. The corresponding electron energy density is given by $U'_{\rm e}$ (in [erg\,cm$^{-3}$]). Also, $\gamma'_{\rm min}$, $\gamma'_{\rm br}$, and $\gamma'_{\rm max}$ are defined as the minimum, break, and maximum Lorentz factor, respectively. We stress that a simple power-law model, which has one degree of freedom less, provides a significantly worse description of the X-ray spectrum. Hence, such a simple function is not a solution here and a model including a break (or any kind of steepening) is required to describe the X-ray emission. The size of the ``variable'' component is set to $R'=2\times10^{16}$\,cm. It is in agreement with the constraint from causality arguments, $R' \lesssim \delta \cdot c \cdot t_{\rm var, obs}$ \citep{1998ApJ...509..608T}, where $t_{\rm var, obs}$ is the observed variability timescale, being $t_{\rm var, obs}\sim1$\,day in the X-ray and VHE (see Sect.~\ref{sec:mwl_variability_sec}). As described in the previous paragraph, $\delta$ is fixed to 10.\par 

Regarding the ``core'' component, the electron distribution is modelled with a simple power-law function, $dN'/d\gamma' = N'_0  \gamma'^{-n}$ with $\gamma'_{\rm min}<\gamma'<\gamma'_{\rm max}$, since the sparsity of the data between the radio and optical/UV does not allow us to constrain functions with a higher degree of complexity. To limit the degrees of freedom, we set $\gamma'_{\rm max}$ to the value at which the synchrotron cooling break occurs. For this, we equate the synchrotron cooling time to the advection time (or effective escape time) of the electrons given by $t'_{\rm esc} \approx R'/c$ \citep{1996ApJ...463..555I}.\par

The ``core'' component from our model is considered as the radio core of 1ES~2344+514 unveiled by VLBI data. Hence, the radius is fixed to $R'=10^{17}$\,cm, similar to the size of the radio core at 15.4\,GHz. Based on VLBI data, \citet{2013A&A...556A..67A} reported a size for the 15.4\,GHz core of $0.07\pm0.04$\,mas, equivalent to $\sim 2 \times 10^{17}$\,cm at the distance of 1ES~2344+514. As an additional constraint, we require the ``core'' component to be at equipartition --- the electron energy density is equal to that of the magnetic field, $U'_{\rm e}=U'_{\rm B}$. Between the two epochs modelled here, the physical parameters of the ``core'' component are identical given the low flux variability observed in the radio (see Sect.~\ref{MWL_variability}).\par 

The resulting models are shown in Fig.~\ref{deep_exposures_nights_sed}: the dark-blue dashed line is the emission from the ``core'', the violet dash-dotted line is the emission from the ``variable'' region, and the light-blue solid line is the sum of both components. The dark dotted line, around $\sim 10^{14}$--$10^{15}$\,Hz, is the contribution from the host galaxy estimated with an elliptical galaxy template borrowed from SWIRE database and scaled to the redshift of 1ES~2344+514. The values of the model parameters derived for both days and both emitting components are listed in Table~\ref{tab:ssc_parameters_deep_exposures}. The model is able to describe the data from radio to VHE satisfactorily. The ``core'' component only dominates in the radio, and is clearly negligible in the gamma rays with respect to the ``variable'' region.\par 

By construction and as mentioned previously, the parameters of the ``core'' component are assumed to be the same between the two days. This assumption seems to be reasonable as it provides an appropriate prediction of the radio flux (at 15\,GHz as measured by OVRO). Regarding the ``variable'' region, the most significant differences in the parameters between the ``deep exposure 1'' and ``deep exposure 2'' models are related to the electron population. More specifically, the high-energy slope of the electron distribution (given by $n_2$) is softer during ``deep exposure 2''. Additionally, $\gamma'_{br}$ is slightly lower on the latter day: this is consistent with the ``harder-when-brighter'' trend reported in Sect.~\ref{sec:X_ray_spectral_deep_exposures}.\par 

The break in the electron distribution in the ``variable'' region for the ``deep exposure 1'' SED is $n_2-n_1 = 1.06$ and occurs at $\gamma'_{br} = 4.9 \times 10^{5}$. Homogeneous models (as the one considered here) predict an electron cooling break due to synchrotron radiation of $n_2-n_1 = 1$, similar to the value obtained here. Furthermore, assuming an electron advection time of $t'_{\rm esc} = R'/c$ \citep[see, e.g.,][]{1996ApJ...463..555I}, the theoretical break location falls within $\sim25\%$ from the value of our model. We stress that the location and the strength of the break are derived directly from X-ray data and not fixed \textit{a priori} based on physical assumptions. The break observed during the ``deep exposure 1'' epoch may thus be associated with synchrotron cooling. For ``deep exposure 2'', although $\gamma'_{br}$ agrees well with synchrotron cooling (within $\sim25\%$), its strength is greater ($n_2-n_1 \approx 1.8$) than theoretically expected if caused solely by synchrotron radiation. Either additional effects related to the source geometry \citep[for instance, inhomogeneities in the emitting region; see][]{2009ApJ...703..662R} are taking place, or the break may also be intrinsic to the underlying acceleration process.\par 

The extrapolation to lower energies of the X-ray emission from the ``variable'' region falls below the UV data points, leading to a similar ``UV excess'' reported in Sect.~\ref{sec:X-ray_spectral_study}. In our model, the ``UV excess'' is due to a significant contribution by the ``core'' region to the UV flux. The additional ``core'' component contributes to the total UV flux at the level of $45\%$ for both epochs.\par 

As an alternative to the broken power-law function to model the electron distribution, we test a log-parabolic model with a low-energy power-law branch \citep[LPPL;][]{2006A&A...448..861M}: 
\begin{equation}
    \label{LPPL_function}
    \frac{dN'}{d\gamma'}(\gamma')= \begin{cases}
    N'_0\, \gamma'{}^{-n}, \quad \gamma'_{\rm min}<\gamma'<\gamma'_{c}\\
    N'_0\, \gamma'{}^{-n - r \log{\gamma'/\gamma'_c}}, \quad \gamma'_{c}<\gamma'<\gamma'_{\rm max}\, ,\\
    \end{cases}
\end{equation}
\noindent
where $r$ quantifies the curvature, and $\gamma'_c$ is the energy above which the curvature occurs in the distribution. The motivation for testing this alternative function is twofold. First, several works showed that a log-parabola distribution may naturally arise in case of stochastic acceleration in magnetic turbulence or any acceleration process in which the acceleration probability is energy dependent \citep{1962SvA.....6..317K, 2004A&A...422..103M, 2006A&A...448..861M, 2009A&A...501..879T}. The power-law branch in the low-energy part of the distribution may originate from first-order Fermi acceleration (diffusive shock acceleration). The LPPL function is thus physically motivated by simultaneous Fermi first-order and second-order (i.e., stochastic) acceleration processes. Second, as shown in Sect.~\ref{sec:X_ray_spectral_deep_exposures}, the sensitive X-ray observations clearly indicate a curved spectrum in agreement with a log-parabola shape.\par 

In order to model the SEDs with Eq.~\ref{LPPL_function}, the slope of the low-energy power-law branch ($n$) is fixed to $n_1$, being the index of the low-energy part of the broken power-law model used previously (see Table~\ref{tab:ssc_parameters_deep_exposures}). Further, $\gamma'_{\rm min}$ and $\gamma'_{\rm max}$ are fixed to the same values as in Table~\ref{tab:ssc_parameters_deep_exposures}. The fitting routines of JetSeT are eventually used to find the best-fit parameters for the electron distribution.\par

Both SEDs are well described with the LPPL function as a model for the electron distribution. The data do not allow us to find a significant preference between the broken power-law and LPPL models. For ``deep exposure 1'', we find $r=1.00$ and $\gamma'_{c}=2.86\times10^5$. Regarding ``deep exposure 2'', we obtain $r=2.17$ and $\gamma'_{c}=2.50\times10^5$. We summarized in Appendix~\ref{low_state_modelling_lppl} the resulting values for all the modelling parameters. From these results, two interesting aspects deserve to be stressed. First, one finds that $r \approx 4 \beta$, where $\beta$ is the curvature of the log-parabola fit to the \textit{XMM-Newton} and \textit{NuSTAR} spectra (see Table~\ref{tab:XMM_nustar_spectral_param}). This relationship is in good agreement with the expected relationship between the synchrotron spectrum curvature and the electron distribution curvature parameter if one assumes the $\delta$-approximation for the computation of the synchrotron flux \citep[see, e.g.,][]{2004A&A...413..489M}. Second, the curvature $r$ is anticorrelated with the location of the synchrotron peak energy: $r$ is twice as high on ``deep exposure 2'' with respect to ``deep exposure 1'', while the synchrotron peak energy is lower ($\nu_{\rm synch,p} = 0.8$\,keV versus $\nu_{\rm synch,p}=1.3$\,keV; see Table~\ref{tab:XMM_nustar_spectral_param}). This anticorrelation trend is consistent with a regime in which stochastic acceleration is the dominant factor responsible for the flux and spectral variability \cite{2011ApJ...739...66T}.

\section{Discussions}
\label{sec:discussion}

\subsection{Intermittent EHBL behaviour}

One of the primary goals of our study is a systematic characterisation of the X-ray and VHE spectral evolution over a multi-year period to better assess the origin of the intermittent EHBL characteristics. Previous works showed that drastic spectral changes implying an EHBL nature can correlate with outburst phases, although the picture remains elusive given that these studies were performed over short time ranges (unlike the one presented here) and also mostly focused on flaring states \citep[e.g.,][]{2000MNRAS.317..743G, 2020MNRAS.496.3912M}.\par 

Using almost three years of \textit{Swift}-XRT data, our results indeed confirm a general ``harder-when-brighter'' evolution in the X-ray with an EHBL behaviour in the synchrotron domain more likely to occur in higher emission states. The anticorrelation between the power-law index $\Gamma_{\rm XRT}$ and the 2--10\,keV flux is significant at the level of $4.6\sigma$ (see Sect.~\ref{sec:X-ray_spectral_study}). Above a 2--10\,keV flux of $\sim 1.7 \times 10^{-11}$\,erg\,cm$^{-2}$\,s$^{-1}$, all spectra display $\Gamma_{\rm XRT}<2$, i.e. a synchrotron peak energy $\geq 1$\,keV and thus behaving as an EHBL. During the  ``deep exposure 1'' and ``deep exposure 2'' epochs, the synchrotron peak frequency is precisely determined thanks to \textit{XMM-Newton}, \textit{NuSTAR}, and \textit{AstroSat} data. A small but significant shift of the peak frequency is indeed visible between these epochs and correlates positively with the flux. In conclusion, this means that flux enhancements are likely accompanied by an injection of freshly accelerated electrons in the radiation zone or due to some reacceleration process \citep[see, for instance,][]{2021A&A...654A..96Z}.\par 

Nonetheless, the unbiased data sample from \textit{Swift}-XRT unveils several outliers to this trend, calling for more complexity. First, one of the \textit{softest} X-ray spectra from this campaign (on October 5\textsuperscript{th} 2019) actually coincides with the highest 0.3--2\,keV flux (by far) and also close to the highest 2--10\,keV flux. This outburst has to be caused by an underlying mechanism that differs from the one driving the general ``harder-when-brighter'' trend visible throughout the campaign. It must be a process that remains achromatic or only slightly chromatic, like a change in the electron density or in the Doppler factor $\delta$. Another interesting outlier is September 20\textsuperscript{th} 2020, where despite the low X-ray flux the \textit{Swift}-XRT index is $\Gamma_{\rm XRT}\approx1.8$, implying a synchrotron peak energy significantly above 1\,keV. Fig.~\ref{synchrotron_remarkable_day} shows a comparison of the corresponding X-ray SED (green markers) with the deep exposures by \textit{XMM-Newton} and \textit{NuSTAR} during other low-activity epochs, which are significantly softer. It is interesting to see in Fig.~\ref{synchrotron_remarkable_day} evidence of an anticorrelation between the soft X-ray ($\sim 10^{17}$\,Hz) and the hard X-ray points ($\sim 10^{18}$\,Hz) when comparing the green (September 20\textsuperscript{th} 2020) and maroon (``deep exposure 2'' -- August 6\textsuperscript{th} 2021) SEDs. This implies relatively strong changes in the acceleration or cooling efficiency of the electrons without a significant change in the flux level.\par

An increase of the acceleration efficiency without a significant change in the electrons injection luminosity in the emitting region could qualitatively explain this behaviour. In such a case, the whole electron distribution would shift to higher energies, thus amplifying the flux at higher energies and decreasing the one at lower energies. Another possibility could be a variation of the magnetic field, while keeping the acceleration efficiency constant: assuming a system close to equilibrium in which synchrotron radiation and adiabatic expansion are the dominant cooling processes, a decrease of $B'$ over time would shift the cooling break in the electron distribution to higher energy, hence leading to an anticorrelation between the rising and falling edges of the synchrotron bump. \citet{2021A&A...655A..89M} put forward such a scenario to explain a $3\sigma$ anticorrelation in Mrk\,421 between the UV and X-ray emission (which correspond to the rising and falling edge of the synchrotron SED component).\par

We note that intermittent shifts of the synchrotron peak above 1\,keV in low activity is particularly rare, and to our knowledge it was only found in Mrk~501 so far \citep{2018A&A...620A.181A}. In conclusion, our results consolidate the hypothesis that being an EHBL is a temporary feature instead of an intrinsic characteristic (at least for a subset of EHBLs).\par

As for the spectral evolution at VHE, the MAGIC data do not allow us to perform a systematic spectral study on a daily timescale. The lower-than-usual VHE state throughout most the campaign, below $4\%$\,Crab Nebula, prevents sufficient photon statistics to build daily SEDs with meaningful constraints on the spectral slope for the majority of the days. It is not possible to firmly identify if shifts of the gamma-ray component to higher energies systematically coincide (or not) with the one observed in the synchrotron part. At least, during the VHE flare in August 2019, the MAGIC spectrum could be determined and the power-law slope shows an indication of ``harder-when-brighter", as observed simultaneously in the X-ray. The hardest spectrum is $\Gamma=1.9\pm0.2$, thus still in agreement with a gamma-ray peak below 1\,TeV. Overall, we do not find indication that the source behaved as an \textit{extreme TeV} EHBL despite significant shift of the synchrotron SED component above 1\,keV. From this point of view, 1ES~2344+514 differs from 1ES~0229+200 or 1ES~1426+426, which have similar synchrotron peak frequencies but a gamma-ray component peaking above 1\,TeV \citep{2019MNRAS.486.1741F, 2020NatAs...4..124B}.\par

\subsection{Origin of the ``UV excess'' and modelling of the quiescent state}

In this work, we report strong evidence of (at least) two emitting components contributing to the synchrotron flux from the IR to the X-ray. It is suggested from the detection of a ``UV excess'' during a ``hard low state'' in the X-ray on September 20\textsuperscript{th} 2020: the UV flux as measured by \textit{Swift}-UVOT lies significantly above the extrapolation to lower energies of the \textit{Swift}-XRT spectrum. The SED modelling of the two long exposures with sensitive X-ray data from \textit{XMM-Newton} and \textit{NuSTAR} (July 23\textsuperscript{rd} 2020 and August 6\textsuperscript{th} 2021) also indicates the presence of such an excess, as discussed earlier. The excess is especially apparent on September 20\textsuperscript{th} 2020 with respect to the other days, which is likely due to the combination of a low, but hard X-ray spectrum on this particular date.\par

A discontinuity in the low-energy SED component was previously noted in the low-frequency peak BL Lac (LBL) object AO~0235+16 \citep{2006A&A...452..845R}. The UV-to-X-ray spectrum of AO~0235+16 may be reconciled by introducing a thermal component. The latter hypothesis is, however, highly improbable for 1ES~2344+514 (see below). Regarding HBLs and EHBLs, some evidence was found in  1ES~0229+200, RGB~0710+591 \citep{2018MNRAS.477.4257C}, and Mrk~501 \citep{2020A&A...637A..86M} based on SED modelling. Nevertheless, the UV/X-ray mismatch shown here for 1ES~2344+514 is quite pronounced and evident: the energy flux around the low-energy end of the hard \textit{Swift}-XRT spectrum on September 20\textsuperscript{th} 2020 is equal to or even below the one in the \textit{Swift}-UVOT filters, which was not the case in the studies mentioned before.\par 

The host galaxy is unlikely to be the origin of the ``UV excess'' given that in the UV the contribution is negligible (in particular in the $UVM2$ and $UVW2$ filters). As also discussed by \citet{2018MNRAS.477.4257C}, an unaccounted UV thermal component from the host (e.g., because of a burst of massive star formation following a recent merger) is very unlikely since this emission would need to be more than an order of magnitude larger than the template used in this work to model the thermal emission from a giant elliptical galaxy. Instead, a more natural solution is that this excess originates from a second population of nonthermal electrons. Such a second population might be less energetic than the one emitting the X-ray/gamma-ray flux and located in larger regions, possibly in downstream parts of the jet.\par

To investigate the latter scenario, we use the two MAGIC observations simultaneous with \textit{NuSTAR}, \textit{XMM-Newton}, and \textit{AstroSat} (July 23\textsuperscript{rd} 2020 and August 6\textsuperscript{th} 2021), which provide the most detailed broad-band view of the quiescent activity of this source (see Sect.~\ref{low_state_model}). We find that a two-component scenario like the one mentioned above is able to well describe the SEDs. In our model, the IR/optical to VHE flux is dominated by a relatively compact region (dubbed as ``variable'') filled with highly energetic electrons ($\gamma'_{\rm min}=2\times10^3$ \& $\gamma'_{\rm max}=8\times10^6$). The second component (dubbed as ``core'') is then introduced to explain the ``UV excess''. The corresponding electron population is less energetic than in the ``variable'' region ($\gamma'_{\rm min}=10$ \& $\gamma'_{\rm max}=9\times10^4$). Its synchrotron SED peaks in the UV band, while in the gamma-ray band it is subdominant by more than an order of magnitude with respect to the ``variable'' region.\par  

Further, our modelling supports the idea that this second component responsible for the ``UV excess'' can be attributed to the $10$\,GHz radio core. As seen in Fig.~\ref{deep_exposures_nights_sed}, it well explains the flux measured by OVRO at 15\,GHz, and the region radius ($\sim10^{-1}$\,pc) is in agreement with VLBI observations at similar frequencies. We also note that a break happens in the model below $\sim10^{11}$\,Hz (see Fig.~\ref{deep_exposures_nights_sed}), which is due to synchrotron self-absorption \citep{1981ApJ...243..700K}. Based on the parameters of the ``core'' component, we find that the self-absorption optical depth $\tau_\nu$ becomes $\sim 1$ at a frequency of about 5\,GHz. Since the surface of unity optical depth is a defining property of a radio core \citep{1979ApJ...232...34B}, this further suggests that the ``UV excess'' comes from the radio ``core'' at a frequency close to 10\,GHz.\par 

\citet{2016A&A...593A..98L} presented a study of the optical ($R$-band) and radio (15\,GHz data from OVRO) emission from a sample of VHE-emitting BL Lac type objects. By exploiting the variability patterns, \citet{2016A&A...593A..98L} extracted information about the different regions contributing to the jet emission. In 1ES~2344+514, it was concluded that at least $25\%$ of the optical flux is radiated in the same region as the one responsible for the 15\,GHz emission. The model we present in Sect.~\ref{low_state_model} is in good agreement with this conclusion as it predicts that $\sim 40\%$ of the total (host-galaxy subtracted) $R$-band flux comes from the region describing the radio data.\par 

In Sect.~\ref{sec:radio_he_correlation_study}, we report a $3\sigma$ positive correlation between OVRO and \textit{Fermi}-LAT. A $2\sigma$ indication was reported by \citet{2018MNRAS.480.5517L}, also using OVRO and \textit{Fermi}-LAT, but over a $\sim 9$\,yr period. In this work, using a more extensive set over $\sim 13$\,yr, the significance is marginally higher. The highest correlation occurs at a time lag of $\sim 90$\,days (meaning the radio lags behind the gamma rays), but the uncertainty is large and the lag is not precisely constrained. Future long-term monitoring is needed to confirm the existence of a correlation and to obtain a better constraint on the time lag that is needed to narrow down the underling physical process. A positive correlation with the radio lagging behind the gamma rays is relatively common in BL Lacs \citep{2014MNRAS.445..428M}. One plausible explanation is the evolution of the electron population, which first emits the gamma-ray flux and then cools down to lower energies to radiate radio photons. The cooling time thus drives the radio/gamma-ray lag, which can reach monthly timescales \citep[see, e.g.,][]{2015MNRAS.448.3121H}. In an alternative scenario, the lag is explained by an evolution of the optical transparency throughout the jet. The gamma-ray emitting zone starts opaque at radio frequencies (because of synchrotron self-absorption), and due to (for example) adiabatic expansion \citep[for detailed modelling, see][]{2022A&A...658A.173T}, gradually becomes transparent to the radio. In the framework of our modelling for the low state presented above, the former scenario is favoured. The high $\gamma'_{\rm min}$ in the ``variable'' region ($\sim 10^3$) implies that the emission in the 15\,GHz band is essentially negligible compared to the ``core'' region and, thus, the radio opacity does not play a role.


\subsection{Time-dependent modelling of the 2019 flaring activity}
\label{flare_modelling}

The VHE flare that occurred in August 2019 reveals complex emission patterns, as discussed in Sect.~\ref{sec:X-ray_vhe_correlation_study}. The VHE flux shows a hint of correlated variability with the 2--10\,keV flux, but not with the 0.3--2\,keV emission, which is stable throughout the flare despite a decay of the VHE flux by a factor of $\sim 3.5$ (see Fig.~\ref{VHE_vs_swift}). The VHE flare appears orphan when one considers only the 0.3--2\,keV energies. In the other bands, no outburst is measured in the UV, optical, and radio. However, the \textit{Fermi}-LAT measurements unveil an approximately daily timescale flare in the 0.3--300\,GeV band coinciding with the VHE. Below, we argue that such multiwavelength variability features are at odds with a one-zone SSC scenario, and two emitting components (at least) must be invoked to describe the multiwavelength behaviour in a leptonic approach.\par 

First, the absence of significant variability in the 0.3--2\,keV regime naturally points toward a corresponding emitting region different than the one responsible for the gamma-ray flare. The indication of variability in the 2--10\,keV band (by a factor $\sim 2$) may be reconciled by considering that the flaring region remains subdominant in the soft X-ray (0.3--2\,keV), but brings a sizable contribution to the 2--10\,keV flux. In principle, changes affecting \textit{only} the highest energy part of the electron distribution could induce a variability of the synchrotron flux in the 2--10\,keV band but not in the 0.3--2\,keV band. However, this scenario would have issues for explaining a \textit{Fermi}-LAT flare (that we report here) given that the electrons radiating MeV--GeV photons (via IC scattering) emit synchrotron photons at much lower energies than $\sim 1$\,keV. \citet{2005A&A...433..479K} investigated the VHE versus X-ray correlation behaviour in (TeV-emitting) BL Lac type objects using a one-zone SSC model. By evolving different physical parameters driving the flux variability, the authors came to the conclusion that the VHE/X-ray correlation is mostly linear or quadratic. Although the exact correlated behaviour is parameter-dependent and sensitive to the precise energy ranges under consideration, any correlation above a quadratic relationship, as we clearly see here between the $>300$\,GeV and 0.3--2\,keV emission, demands either an unphysical scenario or a strong fine-tuning of the parameters. Finally, the evidence of a \textit{Fermi}-LAT flare but not in the UV/optical also points toward (at least) two emitting regions since those SED regions are expected to originate from the same electron population within typical one-zone SSC models applied to TeV BL Lacs \citep{1998ApJ...509..608T}. In summary, within a leptonic scenario, it is highly improbable that the SED from optical/UV to $>1$\,TeV is dominated by a single component during the flare.\par

Following the above arguments, we attempt to describe the flare using a time-dependent leptonic model involving several emitting components. We carry out the modelling over the three-days time span during which we gathered simultaneous MAGIC and \textit{Swift} data (August 6\textsuperscript{th}, 7\textsuperscript{th}, and 8\textsuperscript{th} 2019 --- MJD~58701, MJD~58702, and MJD~58703). We consider a ``quiescent'' component describing the emission state before the flare state and a ``flaring'' component responsible for the gamma-ray flare. The ``quiescent'' component is assumed to be in a steady state, while the ``flaring'' component is evolving with time. The JetSeT modelling code is used to evolve the electron distribution in time. In our scenario, the two components are not cospatial, and thus not interacting with each other.\par 

In order to limit the degrees of freedom in the model, the radius $R'$ and the Doppler factor $\delta$ of the ``quiescent'' component are fixed to the values adopted in Sect.~\ref{low_state_model} (i.e., $R'=2\times10^{16}$\, cm and $\delta=10$). As for the electron distribution, a simple power-law model (without break) is used, $dN'/d\gamma' \propto \gamma'^{-n}$ with $\gamma'_{\rm min}<\gamma'<\gamma'_{\rm max}$. No \textit{XMM-Newton} or \textit{NuSTAR} data are available over this time period to precisely constrain the SED below and above the synchrotron peak, which would be needed to constrain a possible break in $dN'/d\gamma'$. The total electron energy density is given as $U'_{\rm e}$.\par

In the ``flaring'' component a power-law distribution of electrons is instantaneously injected at $t_0 = \mathrm{MJD~58701}$ with a slope of $n'_{\rm inj}$ and with Lorentz factors between $\gamma'_{\rm min,inj}$ and $\gamma'_{\rm max,inj}$. The initial energy density is $U'_{\rm e,inj}$. The electrons subsequently cool, leading to the decay of the flux as observed in the data. We take into account synchrotron, IC, and adiabatic cooling mechanisms. Following \citet{2022A&A...658A.173T}, the adiabatic expansion of the ``flaring'' component occurs at a constant velocity $\beta'_{\rm exp} = v'_{\rm exp}/c$ such that the adiabatic cooling timescale becomes $t'_{\rm ad}=R'(t)/\beta'_{\rm exp}c$. We define $R'_0 = R'(t_0)$ as the initial radius of the ``flaring'' component.  Owing to energy conservation \citep{RevModPhys.56.255}, the magnetic field depends on $R'(t)$ as $B'(t) = B'_0 \left(R'_0/R'(t)\right)^{m_B}$, where $B'_0$ is the initial magnetic field strength. We fix here $m_B = 1$, implying a fully toroidal configuration of the magnetic field \citep{RevModPhys.56.255}. This assumption is in agreement with the results of \citet{2022A&A...658A.173T}, who performed a long-term time-dependent modelling of TeV HBL objects. We assume a value of $\delta=20$ for the Doppler factor, twice the one of the ``quiescent'' component but still within the standard range of values found in modelling TeV BL Lac objects \citep{2010MNRAS.401.1570T}.\par 

\begin{figure}[h!]
   \centering
   \includegraphics[width=1.\columnwidth]{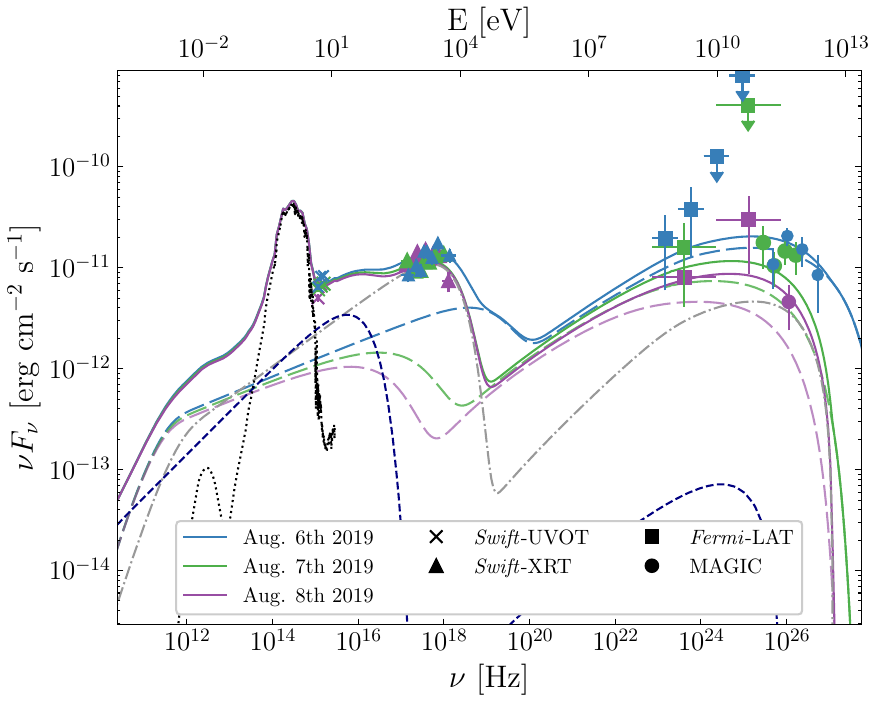}
   \caption{Leptonic modelling of the August 2019 flare. The dash-dotted grey line is the emission from the ``quiescent'' component adopting the parameters listed in Table~\ref{tab:SED_model_flare_quiescent_param}. The long dash line is the emission from the ``flaring'' component, using the parameters listed in Table~\ref{tab:SED_model_flare_flaring_param}. While the ``quiescent'' component is assumed to be constant over time, the ``flaring'' component evolves with time (see text for more details). The short dash blue line depicts the emission from the ``core'' component using the same parameters as the one obtained in Sect.~\ref{low_state_model}, Table~\ref{tab:ssc_parameters_deep_exposures}. The dotted black line is the host-galaxy contribution, using a template from the SWIRE database \citep{2007ApJ...663...81P}. The solid lines are the sum of all components. Each day is plotted with a different color, while the data from the respective instruments are displayed with distinct markers according to the legend.}
    \label{flare_model}
\end{figure}

The adopted parameter values for the ``quiescent'' and ``flaring'' components are listed in Table~\ref{tab:SED_model_flare_quiescent_param} and Table~\ref{tab:SED_model_flare_flaring_param}. The resulting model is plotted in Fig.~\ref{flare_model}. The emission from the ``flaring'' component is plotted in a dashed line for each day. The emission from the ``quiescent'' component is shown with a grey dotted-dash line. Finally, the contribution from the ``core'' region introduced in Sect.~\ref{low_state_model} is also added, using the same parameters as the one obtained in Sect.~\ref{low_state_model}, Table~\ref{tab:ssc_parameters_deep_exposures}. The sum of all components is given with continuous solid lines. We plot with different colors the daily simultaneous SEDs. Regarding \textit{Fermi}-LAT, the SEDs are averaged over two days for August 6\textsuperscript{th} and August 7\textsuperscript{th} 2019. For August 8\textsuperscript{th} 2019, a two-week integration period is adopted given the absence of significant signal over several days around that day.\par

\begin{table}[h!]
\caption{\label{tab:SED_model_flare_quiescent_param} Model parameters of the ``quiescent'' component for the modelling of the August 2019 flare.} 
\centering
\begin{tabular}{ l c c }     
\hline\hline 
 Parameter & Value \\  
\hline\hline   
$B'$ [G] & $4.6 \times 10^{-2}$ \\
$R'$ [cm] & $2\times10^{16}$ \\
$\delta$ & 10 \\
$U'_{\rm e}$ [erg\,cm$^{-3}$] & $8.2\times10^{-3}$ \\
$n$ & 2.38 \\
$\gamma'_{\rm min}$ & $5 \times 10^{2}$ \\
$\gamma'_{\rm max}$ & $10^{6}$ \\
\hline
\end{tabular}
\tablefoot{\centering See text for a description of the parameters.} 
\end{table}

\begin{table}[h!]
\caption{\label{tab:SED_model_flare_flaring_param} Model parameters of the ``flaring'' component for the modelling of the August 2019 flare.} 
\centering
\begin{tabular}{ l c c }     
\hline\hline 
 Parameter & Value \\  
\hline\hline   
$B'_0$ [G] & $10^{-2}$ \\
$R'_0$ [cm] & $7\times10^{15}$ \\
$\beta'_{\rm exp}$ & $3 \times 10^{-3}$ \\
$\delta$ & 20 \\
$U'_{\rm e,inj}$ [erg cm$^{-3}$] & 0.23 \\
$n_{\rm inj}$ & 2.65 \\
$\gamma'_{\rm min, inj}$ & $5 \times 10^{2}$ \\
$\gamma'_{\rm max, inj}$ & $10^{7}$ \\
\hline
\end{tabular}
\tablefoot{\centering See text for a description of the parameters.} 
\end{table}

In general, a relatively good description of the data is achieved from X-ray to TeV. All data points lie within $\lesssim2\sigma$ from the modelling curves. The UV data (from \textit{Swift}-UVOT) is explained by the emission from the ``core'' and ``quiescent'' component, in analogy to the modelling performed in Sect.~\ref{low_state_model}.\par 

Given the multiwavelength coverage and temporal sampling, it is challenging to obtain a precise constraint on the adiabatic expansion velocity $\beta'_{\rm exp}$. Nonetheless, it is interesting to remark that the adopted value, $\beta'_{\rm exp}=3 \times 10^{-3}$, is within the $1\sigma$ uncertainty band obtained by \citet{2022A&A...658A.173T}, who modelled long-term light curves of the HBL Mrk~421 using a time-dependent SSC scenario. In our model, values significantly larger, $\beta'_{\rm exp}\gtrsim 10^{-2}$, are excluded as this leads to a decay of the overall IC flux that would be faster than the observed daily timescale variability. The adiabatic expansion gives rise to a drop of the IC flux that is particularly pronounced owing to a combination of the increasing electron volume and reduction of the target photon density.\par

The ``flaring'' component is significantly Compton dominated. Relative to the ``quiescent'' zone, it remains subdominant in the X-ray up to $\sim 10$\,keV but it dominates in the gamma-ray regime, particularly during the first two days (August 6\textsuperscript{th} and August 7\textsuperscript{th} 2019). The corresponding \textit{Compton dominance}, defined as $A_C = L_{\rm IC, peak}/L_{\rm synch, peak}$, where $L_{\rm IC, peak}$ is the luminosity at the IC peak and $L_{\rm synch, peak}$ is the one at the synchrotron peak, is around $A_C \approx 4$--5 throughout the three days modelled here. Such a high $A_C$ is achieved by invoking an emitting region that is more compact and with a higher electron energy density than the ``quiescent zone". The strong \textit{Compton dominance} leads to an important impact of the IC cooling effects (taken into account in our model), which is even the dominant timescale for electrons with Lorentz factors $\gamma' \approx 10^5$--$10^6$. Regarding the ``quiescent'' component, we have $A_C \approx 0.5$, typical of BL Lacs type objects in nonflaring states \citep{2013ApJ...763..134F}.\par 

The injected electron distribution in the ``flaring'' component is softer ($n_{\rm inj}=2.65$) with respect to the one in the ``quiescent'' zone ($n=2.38$). $n_{\rm inj}$ is somewhat constrained by the \textit{Fermi}-LAT observations since a harder injected spectrum (for instance, with $n_{\rm inj}<2.5$) would start to underpredict the \textit{Fermi}-LAT measurements. Those derived indices are somewhat larger than the canonical index of $\sim 2.23$ expected in shock acceleration \citep{2000ApJ...542..235K}. On the other hand, in the case of oblique shocks (as in a recollimation scenario, for instance), the index of the particle distribution may well be larger, $>2.5$ \citep[see][and references therein]{10.1093/mnras/stw2344}. Interestingly, \citet{2021A&A...654A..96Z} showed that oblique shocks are able to reproduce the SED of EHBLs. Electron acceleration may also be caused in magnetic reconnection events \citep{2014ApJ...783L..21S}, although the relatively low magnetisation of the ``flaring'' component in the model ($B' \approx 10^{-2}$\,G) tends to disfavour such an acceleration process.

   

\section{Concluding remarks}
\label{sec:conclusion}

This paper presents an extensive multiwavelength characterisation of 1ES~2344+514 over a 3\,yr period from radio to VHE. The source is known for its strong spectral variability in the X-ray band and shows intermittent EHBL characteristics. While being among the first extragalactic sources detected at VHE, and also among the closest BL Lac objects, the published multiwavelength campaigns only focus either on flaring states or on small timescales (of a few months). We have organised a VHE monitoring by MAGIC, accompanied especially by simultaneous \textit{Swift}, \textit{XMM-Newton}, \textit{NuSTAR}, and \textit{AstroSat} observations, in order to obtain a systematic and detailed broad-band characterisation of the intermittent EHBL properties. The gathered observations also allowed to study with unprecedented accuracy the low-emission state.\par 

Our results confirm the previously ``harder-when-brighter'' trend in the X-ray band, with a higher probability to find the source in an EHBL state during enhanced flux periods. Nonetheless, several nights also exhibit EHBL-like characteristics during low activity epochs, indicating strong shift of the synchrotron component independently from the source activity. These results likely imply significant changes of the electron acceleration efficiency, orphan from a significant change in the electron injection luminosity. We also observed an X-ray flaring event (one of the brightest for 1ES~2344+514) which is not characterised by an EHBL-like state, in clear contradiction with the ``harder-when-brighter'' trend. In turn, several flaring mechanisms must coexist, some being chromatic, others being achromatic.\par 

During a hard but low X-ray emission state, we find that the UV flux is above the extrapolation to lower energies of the \textit{Swift}-XRT data. This ``UV excess'' cannot be caused by a thermal component and must be of nonthermal origin. Likely, two separate electron populations (at least) significantly contribute to the synchrotron SED. Using a two-component leptonic model applied to the broad-band SEDs with simultaneous MAGIC, \textit{XMM-Newton}, \textit{NuSTAR}, and \textit{AstroSat} observations, we propose that this ``UV excess'' could originate from the $\sim10$\,GHz radio core.\par  

We do not find significant ($>5\sigma$) VHE versus X-ray correlation along the campaign. During a bright VHE flare in August 2019, the $>300$\,GeV flux shows a hint of correlation with the 2--10\,keV band, while nothing is apparent in the 0.3--2\,keV range. In fact, the 0.3--2\,keV flux is constant despite a decay of the VHE flux by more than a factor 3. We argue that this behaviour is not in agreement with a single-zone leptonic scenario. Instead, using a time-dependent modelling approach, we investigate an alternative multi-zone model in which the flare is caused by a very compact emitting region that remains subdominant in the X-ray (in particular below $\sim 2$\,keV) but dominates the VHE emission. This scenario is in agreement with the flare properties, although it requires a flaring region heavily out of equipartition between the electron and magnetic energy densities.

%
%

\begin{acknowledgements}

A.~Arbet Engels: project leadership, paper drafting and editing, coordination of multiwavelength observations and analysis, MAGIC and \textit{Fermi}-LAT data analysis, variability analysis, theoretical modelling and interpretation;
R.~Chatterjee: \textit{AstroSat} data analysis;
F.~D'Ammando: \textit{XMM-Newton} and \textit{NuSTAR} data analysis, paper drafting and editing;
S.~Das: \textit{AstroSat} data analysis;
D.~Dorner: organisation of multiwavelength observations;
A.~V.~Filippenko: KAIT data analysis, paper editing;
S.~Fukami: paper editing, theoretical modelling and interpretation;
T.~Hovatta: OVRO data analysis;
S.~Kiehlmann: OVRO data analysis;
C.~Leto: \textit{Swift}-UVOT data analysis;
P.~Majumdar: organisation of multiwavelength observations;
M.~Manganaro: organisation of multiwavelength observations;
H.~A.~Mondal: paper drafting and editing, MAGIC and \textit{AstroSat} data analysis, variability analysis;
M.~Perri: \textit{Swift}-XRT data analysis;
C.~M.~Raiteri: WEBT campaign management and data analysis, paper editing;
A.~C.~S.~Readhead: OVRO data analysis;
F.~Verrecchia: \textit{Swift}-UVOT data analysis;
M.~Villata: WEBT campaign management and data analysis;
W.~Zheng: KAIT data acquisition and analysis.
The rest of the authors have contributed in one or several of the following ways: design, construction, maintenance, and operation of the instrument(s); preparation and/or evaluation of the observation proposals; data acquisition, processing, calibration and/or reduction; production of analysis tools and/or related Monte Carlo simulations; discussion and approval of the contents of the draft.\\

We would like to thank the Instituto de Astrof\'{\i}sica de Canarias for the excellent working conditions at the Observatorio del Roque de los Muchachos in La Palma. The financial support of the German BMBF, MPG and HGF; the Italian INFN and INAF; the Swiss National Fund SNF; the grants PID2019-104114RB-C31, PID2019-104114RB-C32, PID2019-104114RB-C33, PID2019-105510GB-C31, PID2019-107847RB-C41, PID2019-107847RB-C42, PID2019-107847RB-C44, PID2019-107988GB-C22 funded by the Spanish MCIN/AEI/ 10.13039/501100011033; the Indian Department of Atomic Energy; the Japanese ICRR, the University of Tokyo, JSPS, and MEXT; the Bulgarian Ministry of Education and Science, National RI Roadmap Project DO1-400/18.12.2020 and the Academy of Finland grant \#320045 are gratefully acknowledged. This work has also been supported by Centros de Excelencia ``Severo Ochoa'' y Unidades ``Mar\'{\i}a de Maeztu'' program of the Spanish MCIN/AEI/ 10.13039/501100011033 (SEV-2016-0588, CEX2019-000920-S, CEX2019-000918-M, CEX2021-001131-S, MDM-2015-0509-18-2) and by the CERCA institution of the Generalitat de Catalunya; by the Croatian Science Foundation (HrZZ) Project IP-2016-06-9782 and the University of Rijeka Project uniri-prirod-18-48; by the Deutsche Forschungsgemeinschaft (SFB1491 and SFB876); the Polish Ministry Of Education and Science grant \#2021/WK/08; and by the Brazilian MCTIC, CNPq and FAPERJ.\\ 

A.A.E and D.P acknowledge support from the Deutsche Forschungs gemeinschaft (DFG, German Research Foundation) under Germany’s Excellence Strategy – EXC-2094 – 390783311

This work made use of data from the \textit{NuSTAR} mission, a project led by the California Institute of Technology, managed by the Jet Propulsion Laboratory, and funded by the National Aeronautics and Space Administration (NASA). We thank the \textit{NuSTAR} Operations, Software, and Calibration teams for support with the execution and analysis of these observations. This research has made use of the \textit{NuSTAR} Data Analysis Software (NuSTARDAS) jointly developed by the ASI Science Data Center (ASDC; Italy) and the California Institute of Technology (USA).
This research has also made use of the XRT Data Analysis Software (XRTDAS) developed under the responsibility of the ASI Science Data Center (ASDC), Italy.
This work has used data from the Indian Space Science Data Centre (ISSDC) under the \textit{AstroSat} mission of the Indian Space Research Organisation (ISRO). We acknowledge the POC teams of the SXT and LAXPC instruments for archiving data and providing the necessary software tools. S.D. and R.C. thank ISRO for support under the \textit{AstroSat} archival data utilization program, and IUCAA for their hospitality and usage of their facilities during their stay at different times as part of the university associateship program.\\

The Abastumani team acknowledges financial support by the Shota
Rustaveli NSF of Georgia under contract FR-19-6174. The $R$-band photometric data from the University of Athens Observatory (UOAO) were obtained after utilizing the robotic and remotely controlled instruments at the facilities \citep{2016RMxAC..48...22G}. The research at Boston University was supported in part by NASA Fermi GI grant 80NSSC22K1571 and U.S. National Science Foundation (NSF) grant AST-2108622. This study was based (in part) on observations conducted using the 1.8\,m Perkins Telescope Observatory (PTO) in Arizona (USA), which is owned and operated by Boston University. This research was partially supported by the Bulgarian National Science Fund of the Ministry of Education and Science under grants KP-06-H38/4 (2019), KP-06-KITAJ/2 (2020) and KP-06-H68/4 (2022). The Skinakas Observatory is a collaborative project of the University of Crete, the Foundation for Research and Technology -- Hellas, and the Max-Planck-Institut f\"ur Extraterrestrische Physik. G.D., O.V., M.D.J., and M.S. acknowledge support by the Astronomical station Vidojevica, funding from the Ministry of Science, Technological Development and Innovation of the Republic of Serbia (contract \#451-03-47/2023-01/200002), by the EC through project BELISSIMA (call FP7-REGPOT-2010-5, \#265772), the observing and financial grant support from the Institute of Astronomy and Rozhen NAO BAS through the bilateral SANU-BAN joint research project GAIA ASTROMETRY AND FAST VARIABLE ASTRONOMICAL OBJECTS, and support by the SANU project F-187. This paper used observations made with the IAC-80 telescope operated on the island of Tenerife by the Instituto de Astrof\'{\i}sica de Canarias in the Spanish Observatorio del Teide and also observations made with the LCOGT 0.4\,m telescope network, one of whose nodes is located in the Spanish Observatorio del Teide.


This research has made use of data from the OVRO 40\,m monitoring program \citep{2011ApJS..194...29R}, supported by private funding from the California Institute of Technology and the Max Planck Institute for Radio Astronomy, and by NASA grants NNX08AW31G, NNX11A043G, and NNX14AQ89G and NSF grants AST-0808050 and AST-1109911. S.K. acknowledges support from the European Research Council (ERC) under the European Unions Horizon 2020 research and innovation program under grant agreement \#771282.


KAIT and its ongoing operation were made possible by donations from Sun Microsystems, Inc., the Hewlett-Packard Company, AutoScope Corporation, the Lick Observatory, the U.S. NSF, the University of California, the Sylvia \& Jim Katzman Foundation, and the TABASGO Foundation. Research at Lick Observatory is partially supported by a generous gift from Google. A.V.F.'s group at U.C. Berkeley is grateful for financial assistance from the Christopher R. Redlich Fund, Alan Eustace (W.Z. is a Eustace Specialist in Astronomy), and many other donors.

\end{acknowledgements}

\bibliographystyle{aa}
\bibliography{bibliography_paper}

\begin{appendix}

\section{\textit{Fermi}-LAT and OVRO long-term light curves}
\label{radio_fermi_longterm_lc}

\begin{figure*}[h!]
   \centering
   \includegraphics[width=2.0\columnwidth]{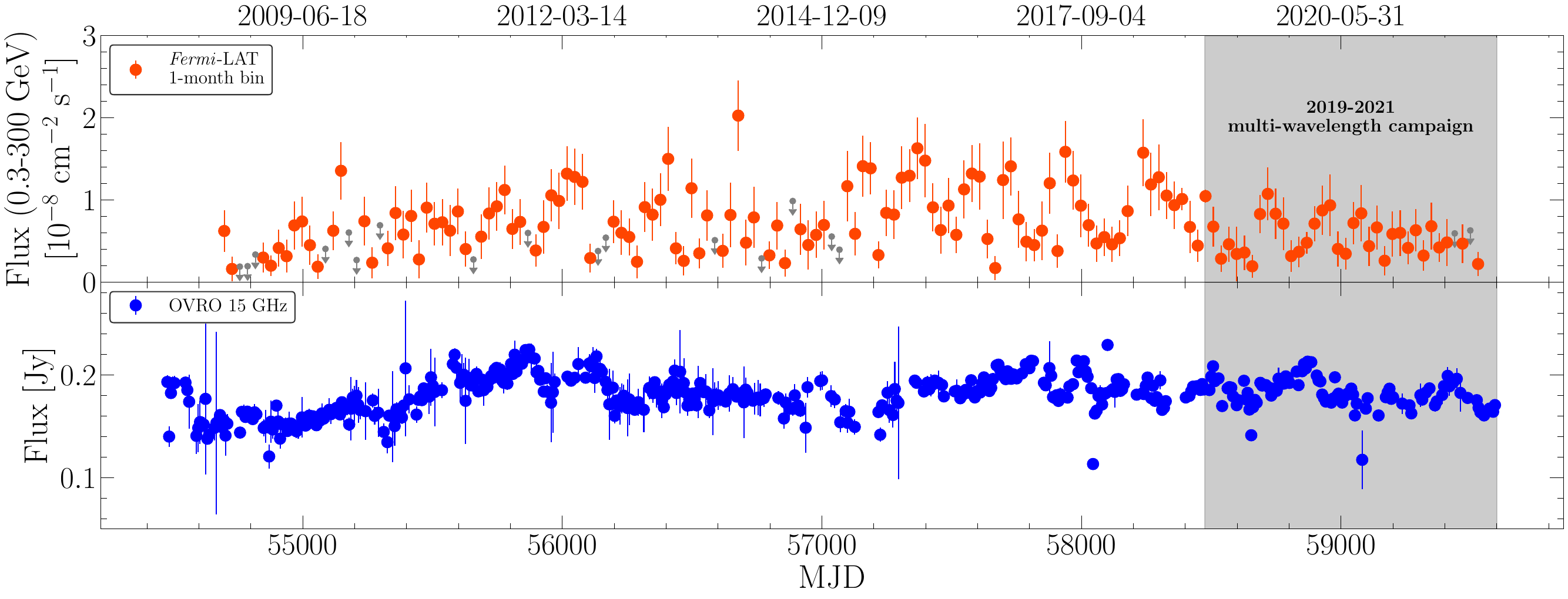}
   \caption{Long-term \textit{Fermi}-LAT (0.3--300\,GeV) and OVRO (15\,GHz) light curves, from 2008 until the end of 2021. The \textit{Fermi}-LAT fluxes are monthly binned. See Sect.~\ref{sec:analysis} and Sect.~\ref{sec:mwl_variability_sec} for more details about the analysis.}
    \label{fermi_radio_lc}
\end{figure*}

This section presents the long-term light curves from \textit{Fermi}-LAT (0.3--300\,GeV) and OVRO (15\,GHz) between 2008 and the end of 2021. The \textit{Fermi}-LAT fluxes are computed with a monthly binning. An upper limit at 95\% confidence level is quoted in case the TS is below 5 (see Sect.~\ref{sec:analysis}). The light curves are shown in Fig.~\ref{fermi_radio_lc}. The grey vertical band highlights the period of time corresponding to the 2019--2021 multiwavelength campaign, which is the focus of this work. Based on these light curves, the correlation between the gamma-ray and radio emission is investigated, and the results are displayed in Sect.~\ref{sec:radio_he_correlation_study}.

\clearpage

\section{\textit{Swift}-XRT spectral fit parameters}\label{sec:swift_xrt_parameters}
This section presents the spectral parameters of all the \textit{Swift}-XRT observations performed throughout the campaign, using power-law and log-parabola models. The models assume a photoelectric absorption by a fixed column density of $N_{\rm H}=1.41\times10^{21}$\,cm$^{-2}$ \citep[][]{2016A&A...594A.116H}. The results are listed in Table~\ref{tab:swift_spectral_param}. 

\begin{table*}[h!]
\caption{\label{tab:swift_spectral_param}\textit{Swift}-XRT analysis results.} 
\tiny
\centering
\begin{tabular}{ l c c c c c c c c c}     
\hline
 MJD & Date & $F_{0.3-2\text{\,keV}}$ & $F_{2-10\text{\,keV}}$ & $\Gamma$ & $\chi^2$/dof & $\alpha$ & $\beta$ & $\chi^2$/dof \\
 & YYYY-MM-DD & $[10^{-11} \mathrm{erg} \, \mathrm{cm}^{-2} \mathrm{s}^{-1}]$ & $[10^{-11} \mathrm{erg} \, \mathrm{cm}^{-2} \mathrm{s}^{-1}]$ & $[10^{-10} \mathrm{erg}\, \mathrm{cm}^{-2} \mathrm{s}^{-1}]$ &  &  &  &  &\\
\hline
58504.8333 & 2019-01-21 & 1.44 $\pm$ 0.07 & 1.08 $\pm$ 0.19 & 2.08 $\pm$ 0.08 & 33/25 & 2.09 $\pm$ 0.12 & -0.03 $\pm$ 0.27 & 33/24\\
58511.5287 & 2019-01-28 & 1.87 $\pm$ 0.11 & 0.96 $\pm$ 0.21 & 2.20 $\pm$ 0.10 & 10/16 & 2.03 $\pm$ 0.17 & 0.55 $\pm$ 0.42 & 8/15\\
58521.9488 & 2019-02-07 & 2.05 $\pm$ 0.10 & 1.88 $\pm$ 0.22 & 1.93 $\pm$ 0.07 & 34/29 & 1.82 $\pm$ 0.12 & 0.26 $\pm$ 0.23 & 33/28\\
58525.6004 & 2019-02-11 & 1.63 $\pm$ 0.10 & 1.63 $\pm$ 0.37 & 1.92 $\pm$ 0.10 & 18/18 & 1.94 $\pm$ 0.16 & -0.07 $\pm$ 0.36 & 18/17\\
58532.0465 & 2019-02-18 & 1.43 $\pm$ 0.06 & 1.27 $\pm$ 0.11 & 1.94 $\pm$ 0.05 & 58/44 & 1.64 $\pm$ 0.11 & 0.63 $\pm$ 0.19 & 45/43\\
58539.8767 & 2019-02-25 & 1.79 $\pm$ 0.09 & 1.87 $\pm$ 0.24 & 1.86 $\pm$ 0.07 & 25/24 & 1.80 $\pm$ 0.14 & 0.16 $\pm$ 0.26 & 25/23\\
58546.2654 & 2019-03-04 & 1.36 $\pm$ 0.06 & 1.30 $\pm$ 0.15 & 1.93 $\pm$ 0.07 & 24/27 & 1.91 $\pm$ 0.13 & 0.05 $\pm$ 0.25 & 24/26\\
58553.3733 & 2019-03-11 & 1.41 $\pm$ 0.06 & 1.09 $\pm$ 0.12 & 2.05 $\pm$ 0.07 & 31/28 & 2.00 $\pm$ 0.12 & 0.11 $\pm$ 0.21 & 31/27\\
58560.0113 & 2019-03-18 & 1.01 $\pm$ 0.05 & 0.90 $\pm$ 0.12 & 1.98 $\pm$ 0.08 & 27/22 & 1.98 $\pm$ 0.14 & -0.01 $\pm$ 0.26 & 27/21\\
58567.3744 & 2019-03-25 & 1.14 $\pm$ 0.08 & 1.06 $\pm$ 0.26 & 1.94 $\pm$ 0.12 & 4/10 & 1.92 $\pm$ 0.20 & 0.05 $\pm$ 0.43 & 4/9\\
58701.0571 & 2019-08-06 & 1.78 $\pm$ 0.12 & 2.25 $\pm$ 0.32 & 1.76 $\pm$ 0.09 & 11/14 & 1.59 $\pm$ 0.20 & 0.33 $\pm$ 0.33 & 10/13\\
58702.1200 & 2019-08-07 & 1.95 $\pm$ 0.11 & 2.05 $\pm$ 0.34 & 1.88 $\pm$ 0.09 & 18/19 & 1.92 $\pm$ 0.16 & -0.08 $\pm$ 0.30 & 18/18\\
58703.1156 & 2019-08-08 & 2.05 $\pm$ 0.12 & 1.43 $\pm$ 0.19 & 2.07 $\pm$ 0.08 & 21/17 & 1.76 $\pm$ 0.17 & 0.70 $\pm$ 0.31 & 15/16\\
58704.7681 & 2019-08-09 & 1.10 $\pm$ 0.07 & 1.46 $\pm$ 0.26 & 1.74 $\pm$ 0.10 & 11/15 & 1.75 $\pm$ 0.20 & -0.01 $\pm$ 0.35 & 11/14\\
58707.9554 & 2019-08-12 & 1.58 $\pm$ 0.11 & 1.14 $\pm$ 0.25 & 2.03 $\pm$ 0.10 & 7/11 & 1.83 $\pm$ 0.20 & 0.52 $\pm$ 0.43 & 6/10\\
58716.1266 & 2019-08-21 & 1.54 $\pm$ 0.09 & 0.99 $\pm$ 0.17 & 2.15 $\pm$ 0.09 & 14/16 & 2.14 $\pm$ 0.15 & 0.05 $\pm$ 0.31 & 14/15\\
58724.0827 & 2019-08-29 & 2.00 $\pm$ 0.07 & 1.71 $\pm$ 0.15 & 1.97 $\pm$ 0.05 & 64/55 & 1.74 $\pm$ 0.10 & 0.48 $\pm$ 0.19 & 56/54\\
58761.2292 & 2019-10-05 & 3.50 $\pm$ 0.09 & 2.13 $\pm$ 0.14 & 2.17 $\pm$ 0.04 & 103/95 & 2.08 $\pm$ 0.06 & 0.22 $\pm$ 0.12 & 100/94\\
59017.1702 & 2020-06-17 & 0.71 $\pm$ 0.05 & 0.69 $\pm$ 0.35 & 1.99 $\pm$ 0.14 & 4/10 & 2.03 $\pm$ 0.19 & -0.19 $\pm$ 0.60 & 4/9\\
59018.1643 & 2020-06-18 & 0.96 $\pm$ 0.06 & 0.82 $\pm$ 0.13 & 1.99 $\pm$ 0.09 & 15/18 & 1.96 $\pm$ 0.15 & 0.08 $\pm$ 0.30 & 15/17\\
59023.1415 & 2020-06-23 & 1.08 $\pm$ 0.06 & 0.95 $\pm$ 0.14 & 1.96 $\pm$ 0.08 & 19/21 & 1.91 $\pm$ 0.14 & 0.14 $\pm$ 0.27 & 19/20\\
59027.1230 & 2020-06-27 & 1.30 $\pm$ 0.07 & 1.23 $\pm$ 0.26 & 2.01 $\pm$ 0.09 & 18/22 & 2.11 $\pm$ 0.13 & -0.34 $\pm$ 0.29 & 16/21\\
59048.1055 & 2020-07-18 & 1.07 $\pm$ 0.04 & 0.75 $\pm$ 0.08 & 2.10 $\pm$ 0.06 & 52/34 & 1.96 $\pm$ 0.12 & 0.30 $\pm$ 0.22 & 50/33\\
59052.0840 & 2020-07-22 & 1.14 $\pm$ 0.04 & 0.82 $\pm$ 0.07 & 2.08 $\pm$ 0.05 & 60/47 & 1.90 $\pm$ 0.10 & 0.40 $\pm$ 0.19 & 55/46\\
59053.0175 & 2020-07-23 & 1.23 $\pm$ 0.03 & 1.03 $\pm$ 0.05 & 1.99 $\pm$ 0.03 & 92/104 & 1.78 $\pm$ 0.06 & 0.45 $\pm$ 0.11 & 73/103\\
59054.2023 & 2020-07-24 & 1.22 $\pm$ 0.05 & 0.93 $\pm$ 0.12 & 2.08 $\pm$ 0.07 & 28/34 & 2.11 $\pm$ 0.10 & -0.10 $\pm$ 0.21 & 28/33\\
59060.0622 & 2020-07-30 & 1.28 $\pm$ 0.06 & 0.87 $\pm$ 0.11 & 2.10 $\pm$ 0.07 & 41/26 & 2.04 $\pm$ 0.11 & 0.17 $\pm$ 0.21 & 41/25\\
59079.1717 & 2020-08-18 & 1.25 $\pm$ 0.06 & 1.08 $\pm$ 0.15 & 1.99 $\pm$ 0.07 & 38/27 & 1.99 $\pm$ 0.12 & 0.00 $\pm$ 0.23 & 38/26\\
59080.4317 & 2020-08-19 & 1.30 $\pm$ 0.07 & 0.96 $\pm$ 0.14 & 2.05 $\pm$ 0.08 & 21/21 & 1.95 $\pm$ 0.13 & 0.25 $\pm$ 0.27 & 21/20\\
59084.1477 & 2020-08-23 & 1.07 $\pm$ 0.04 & 0.79 $\pm$ 0.08 & 2.08 $\pm$ 0.06 & 34/38 & 2.06 $\pm$ 0.09 & 0.04 $\pm$ 0.18 & 34/37\\
59089.1928 & 2020-08-28 & 0.90 $\pm$ 0.04 & 0.82 $\pm$ 0.09 & 1.93 $\pm$ 0.06 & 25/30 & 1.80 $\pm$ 0.12 & 0.30 $\pm$ 0.23 & 23/29\\
59107.1177 & 2020-09-15 & 1.05 $\pm$ 0.04 & 0.71 $\pm$ 0.07 & 2.12 $\pm$ 0.06 & 28/37 & 2.02 $\pm$ 0.10 & 0.23 $\pm$ 0.18 & 26/36\\
59112.4382 & 2020-09-20 & 0.95 $\pm$ 0.04 & 1.06 $\pm$ 0.10 & 1.82 $\pm$ 0.05 & 35/43 & 1.73 $\pm$ 0.10 & 0.19 $\pm$ 0.18 & 34/42\\
59117.0297 & 2020-09-25 & 0.94 $\pm$ 0.04 & 0.90 $\pm$ 0.10 & 1.93 $\pm$ 0.06 & 35/34 & 1.97 $\pm$ 0.11 & -0.07 $\pm$ 0.20 & 34/33\\
59129.3033 & 2020-10-07 & 0.96 $\pm$ 0.04 & 0.74 $\pm$ 0.07 & 2.03 $\pm$ 0.06 & 36/38 & 1.97 $\pm$ 0.09 & 0.19 $\pm$ 0.19 & 35/37\\
59130.9094 & 2020-10-08 & 1.23 $\pm$ 0.05 & 0.91 $\pm$ 0.11 & 2.05 $\pm$ 0.06 & 34/29 & 1.88 $\pm$ 0.13 & 0.38 $\pm$ 0.23 & 31/28\\
59131.2346 & 2020-10-09 & 1.09 $\pm$ 0.06 & 0.80 $\pm$ 0.16 & 2.02 $\pm$ 0.09 & 15/17 & 1.91 $\pm$ 0.15 & 0.35 $\pm$ 0.34 & 14/16\\
59135.0706 & 2020-10-13 & 1.04 $\pm$ 0.06 & 0.72 $\pm$ 0.11 & 2.08 $\pm$ 0.08 & 20/17 & 1.86 $\pm$ 0.16 & 0.53 $\pm$ 0.31 & 17/16\\
59140.0020 & 2020-10-18 & 1.15 $\pm$ 0.05 & 0.84 $\pm$ 0.12 & 2.00 $\pm$ 0.07 & 22/27 & 1.68 $\pm$ 0.15 & 0.80 $\pm$ 0.31 & 15/26\\
59145.0269 & 2020-10-23 & 1.86 $\pm$ 0.07 & 1.19 $\pm$ 0.13 & 2.12 $\pm$ 0.06 & 30/41 & 1.98 $\pm$ 0.10 & 0.38 $\pm$ 0.22 & 27/40\\
59165.7537 & 2020-11-12 & 1.16 $\pm$ 0.04 & 0.97 $\pm$ 0.10 & 1.97 $\pm$ 0.05 & 33/43 & 1.84 $\pm$ 0.10 & 0.32 $\pm$ 0.21 & 30/42\\
59169.2780 & 2020-11-16 & 1.69 $\pm$ 0.10 & 1.25 $\pm$ 0.21 & 2.11 $\pm$ 0.09 & 11/19 & 2.18 $\pm$ 0.13 & -0.21 $\pm$ 0.27 & 11/18\\
59171.8638 & 2020-11-18 & 1.37 $\pm$ 0.05 & 1.30 $\pm$ 0.10 & 1.93 $\pm$ 0.05 & 45/45 & 1.79 $\pm$ 0.10 & 0.26 $\pm$ 0.17 & 42/44\\
59189.1824 & 2020-12-06 & 1.22 $\pm$ 0.05 & 0.81 $\pm$ 0.09 & 2.06 $\pm$ 0.06 & 43/34 & 1.84 $\pm$ 0.11 & 0.62 $\pm$ 0.23 & 34/33\\
59191.1154 & 2020-12-08 & 1.50 $\pm$ 0.09 & 1.04 $\pm$ 0.18 & 2.07 $\pm$ 0.09 & 9/16 & 1.92 $\pm$ 0.16 & 0.39 $\pm$ 0.34 & 7/15\\
59196.9438 & 2020-12-13 & 1.13 $\pm$ 0.05 & 0.73 $\pm$ 0.09 & 2.11 $\pm$ 0.06 & 37/34 & 1.97 $\pm$ 0.11 & 0.38 $\pm$ 0.24 & 35/33\\
59203.8450 & 2020-12-20 & 1.81 $\pm$ 0.09 & 2.14 $\pm$ 0.26 & 1.79 $\pm$ 0.07 & 29/28 & 1.70 $\pm$ 0.14 & 0.19 $\pm$ 0.25 & 28/27\\
59221.8579 & 2021-01-07 & 2.41 $\pm$ 0.08 & 2.27 $\pm$ 0.23 & 1.94 $\pm$ 0.06 & 39/46 & 1.92 $\pm$ 0.10 & 0.04 $\pm$ 0.18 & 39/45\\
59225.8369 & 2021-01-11 & 2.19 $\pm$ 0.08 & 1.90 $\pm$ 0.18 & 1.94 $\pm$ 0.05 & 52/50 & 1.81 $\pm$ 0.09 & 0.33 $\pm$ 0.18 & 48/49\\
59404.0609 & 2021-07-09 & 1.87 $\pm$ 0.06 & 1.62 $\pm$ 0.13 & 1.98 $\pm$ 0.05 & 48/53 & 1.91 $\pm$ 0.08 & 0.15 $\pm$ 0.15 & 46/52\\
59406.1129 & 2021-07-11 & 2.12 $\pm$ 0.11 & 2.17 $\pm$ 0.37 & 1.88 $\pm$ 0.08 & 22/21 & 1.84 $\pm$ 0.14 & 0.09 $\pm$ 0.28 & 22/20\\
59408.1637 & 2021-07-13 & 1.93 $\pm$ 0.07 & 1.60 $\pm$ 0.15 & 1.98 $\pm$ 0.05 & 72/45 & 1.79 $\pm$ 0.09 & 0.43 $\pm$ 0.17 & 64/44\\
59410.0985 & 2021-07-15 & 2.23 $\pm$ 0.08 & 1.69 $\pm$ 0.13 & 2.06 $\pm$ 0.05 & 41/45 & 1.96 $\pm$ 0.10 & 0.22 $\pm$ 0.17 & 39/44\\
59412.0861 & 2021-07-17 & 2.13 $\pm$ 0.07 & 1.85 $\pm$ 0.16 & 1.99 $\pm$ 0.05 & 49/50 & 1.98 $\pm$ 0.09 & 0.01 $\pm$ 0.16 & 49/49\\
59414.0090 & 2021-07-19 & 2.16 $\pm$ 0.08 & 1.97 $\pm$ 0.18 & 1.94 $\pm$ 0.05 & 42/53 & 1.87 $\pm$ 0.09 & 0.17 $\pm$ 0.17 & 41/52\\
59428.0215 & 2021-08-02 & 1.67 $\pm$ 0.08 & 1.08 $\pm$ 0.11 & 2.16 $\pm$ 0.06 & 61/29 & 1.88 $\pm$ 0.14 & 0.56 $\pm$ 0.25 & 55/28\\
59430.0084 & 2021-08-04 & 1.88 $\pm$ 0.10 & 1.31 $\pm$ 0.28 & 2.10 $\pm$ 0.09 & 14/22 & 2.09 $\pm$ 0.14 & 0.06 $\pm$ 0.33 & 14/21\\
59431.0044 & 2021-08-05 & 1.41 $\pm$ 0.05 & 1.13 $\pm$ 0.11 & 2.03 $\pm$ 0.06 & 33/42 & 2.00 $\pm$ 0.10 & 0.07 $\pm$ 0.18 & 33/41\\
59433.0073 & 2021-08-07 & 1.29 $\pm$ 0.06 & 1.13 $\pm$ 0.17 & 1.98 $\pm$ 0.08 & 24/25 & 1.97 $\pm$ 0.13 & 0.03 $\pm$ 0.27 & 24/24\\
59434.5908 & 2021-08-08 & 1.28 $\pm$ 0.04 & 1.11 $\pm$ 0.10 & 1.96 $\pm$ 0.05 & 52/53 & 1.83 $\pm$ 0.09 & 0.30 $\pm$ 0.17 & 49/52\\
59441.0220 & 2021-08-15 & 1.70 $\pm$ 0.06 & 1.22 $\pm$ 0.11 & 2.08 $\pm$ 0.05 & 52/42 & 1.94 $\pm$ 0.10 & 0.30 $\pm$ 0.18 & 49/41\\
59443.0895 & 2021-08-17 & 1.67 $\pm$ 0.07 & 1.86 $\pm$ 0.19 & 1.87 $\pm$ 0.06 & 28/37 & 1.96 $\pm$ 0.10 & -0.20 $\pm$ 0.18 & 27/36\\
59459.0703 & 2021-09-02 & 1.60 $\pm$ 0.12 & 1.89 $\pm$ 0.40 & 1.81 $\pm$ 0.11 & 14/12 & 1.80 $\pm$ 0.22 & 0.02 $\pm$ 0.40 & 13/11\\
59490.5379 & 2021-10-03 & 0.96 $\pm$ 0.04 & 0.80 $\pm$ 0.08 & 1.98 $\pm$ 0.05 & 54/39 & 1.71 $\pm$ 0.12 & 0.57 $\pm$ 0.21 & 45/38\\
59496.9142 & 2021-10-09 & 1.95 $\pm$ 0.07 & 1.45 $\pm$ 0.14 & 2.04 $\pm$ 0.05 & 53/43 & 1.88 $\pm$ 0.10 & 0.39 $\pm$ 0.18 & 48/42\\
59516.2217 & 2021-10-29 & 0.96 $\pm$ 0.06 & 0.70 $\pm$ 0.14 & 2.02 $\pm$ 0.10 & 11/13 & 1.83 $\pm$ 0.19 & 0.50 $\pm$ 0.42 & 9/12\\
59543.1677 & 2021-11-25 & 1.22 $\pm$ 0.08 & 0.94 $\pm$ 0.17 & 2.02 $\pm$ 0.09 & 6/15 & 1.87 $\pm$ 0.17 & 0.36 $\pm$ 0.34 & 5/14\\
59552.0648 & 2021-12-04 & 1.81 $\pm$ 0.11 & 1.87 $\pm$ 0.33 & 1.87 $\pm$ 0.10 & 18/16 & 1.80 $\pm$ 0.18 & 0.15 $\pm$ 0.34 & 18/15\\
59553.0602 & 2021-12-05 & 1.99 $\pm$ 0.10 & 1.60 $\pm$ 0.20 & 1.96 $\pm$ 0.07 & 31/24 & 1.67 $\pm$ 0.15 & 0.71 $\pm$ 0.28 & 23/23\\
59554.1188 & 2021-12-06 & 2.08 $\pm$ 0.12 & 1.45 $\pm$ 0.20 & 2.09 $\pm$ 0.08 & 20/19 & 1.93 $\pm$ 0.16 & 0.36 $\pm$ 0.30 & 19/18\\
59555.1731 & 2021-12-07 & 1.44 $\pm$ 0.10 & 1.26 $\pm$ 0.19 & 1.93 $\pm$ 0.09 & 18/13 & 1.63 $\pm$ 0.22 & 0.66 $\pm$ 0.43 & 15/12\\
59556.1778 & 2021-12-08 & 1.13 $\pm$ 0.06 & 1.01 $\pm$ 0.15 & 1.90 $\pm$ 0.08 & 24/20 & 1.53 $\pm$ 0.18 & 0.82 $\pm$ 0.34 & 17/19\\
\hline 
\end{tabular}
\tablefoot{\tiny For each observation, the 0.3-2\,keV, 2-10\,keV fluxes are given. The best-fit power-law indices $\Gamma$ are listed in the fifth column with the corresponding $\chi^2$/dof in the sixth column. The best-fit parameters $\alpha$ and $\beta$ from the log-parabolic fits with a pivot energy fixed at 1\,keV are also given with their corresponding $\chi^2$/dof.}
\end{table*}

\clearpage

\section{MAGIC spectra for epochs of interest}\label{sec:magic_spectral_fit}

\begin{figure}[h!]
   \centering
   \includegraphics[width=1.0\columnwidth]{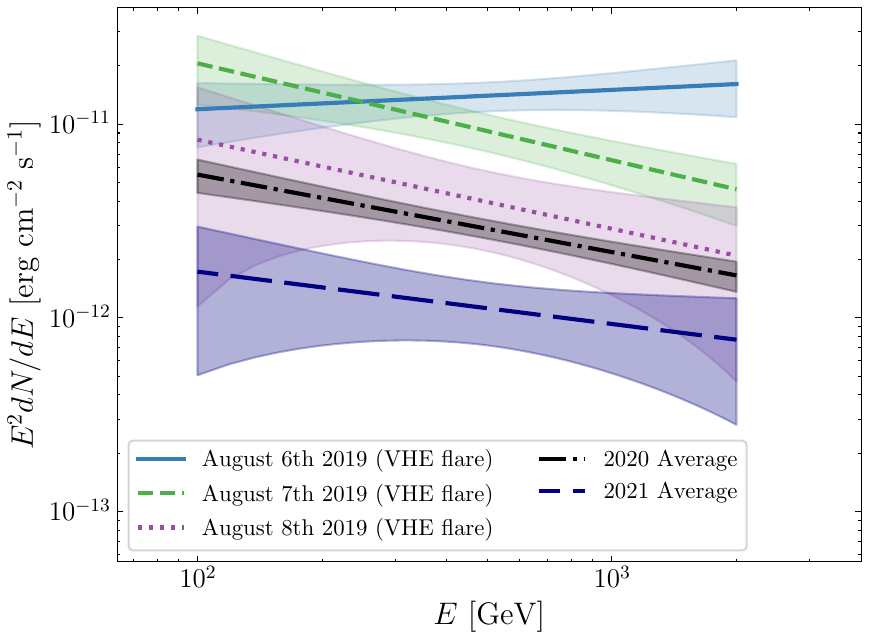}
   \caption{Best-fit power-law models from MAGIC observations during the VHE flare period of August 2019. For comparison purposes, the average spectra of 2020 and 2021 are also shown. More details on the analysis can be found in Sect.~\ref{sec:spectral_study}.}
    \label{magic_seds_flares}
\end{figure}

\begin{figure}[h!]
   \centering
   \includegraphics[width=1.0\columnwidth]{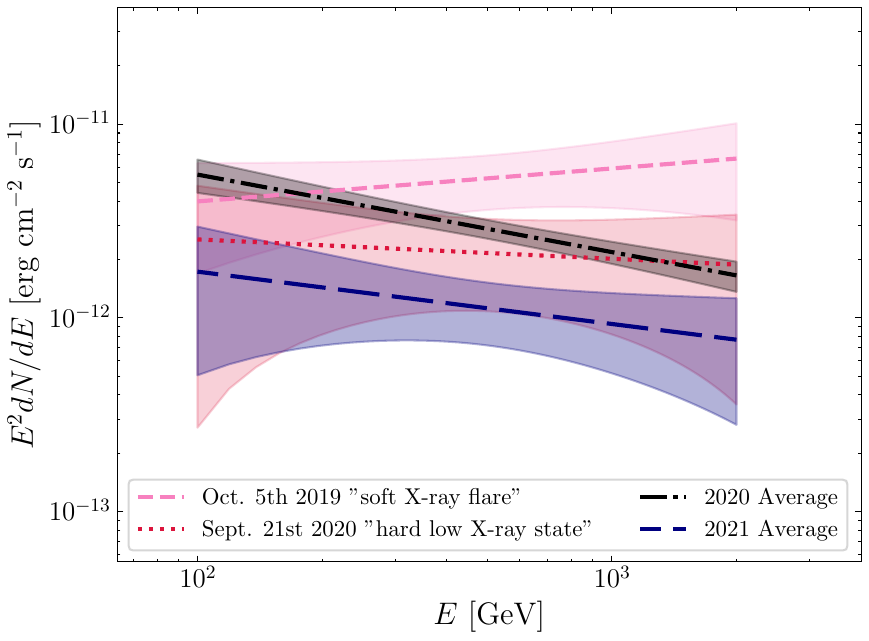}
   \caption{Best-fit power-law models from MAGIC observations during the ``soft X-ray state'' and the ``hard low X-ray state''. For comparison purposes, the average spectra of 2020 and 2021 are also shown. More details on the analysis can be found in Sect.~\ref{sec:spectral_study}.}
    \label{magic_seds_special_days}
\end{figure}

\begin{figure}[h!]
   \centering
   \includegraphics[width=1.0\columnwidth]{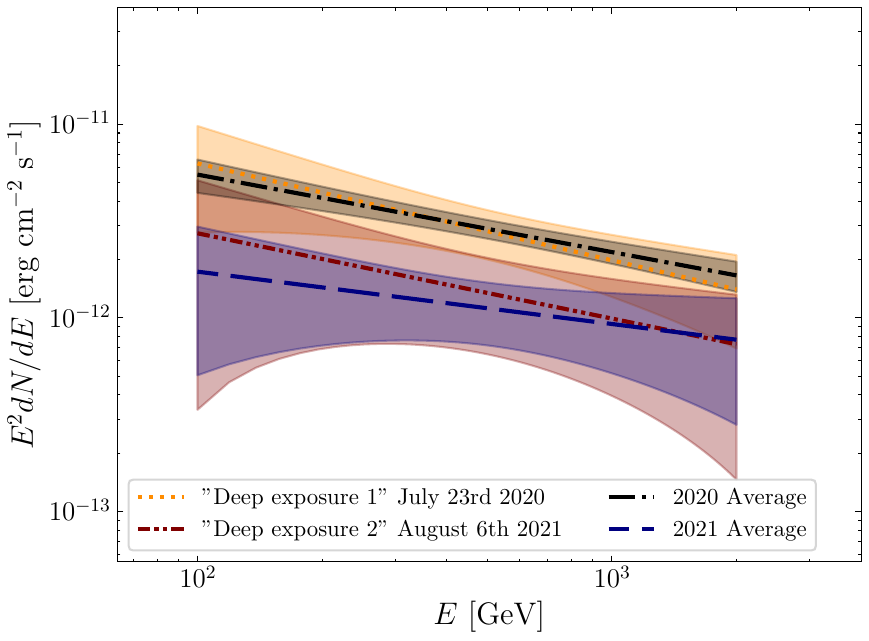}
   \caption{Best-fit power-law models from MAGIC observations during ``deep exposure 1'' and ``deep exposure 2''. For comparison purposes, the average spectra of 2020 and 2021 are also shown. More details on the analysis can be found in Sect.~\ref{sec:spectral_study}.}
    \label{magic_seds_deep_exposures}
\end{figure}

In this section, we show plots of the MAGIC best-fit spectra during the epochs of interest discussed in Sec.~\ref{sec:magic_spectral_study}. As described Sec.~\ref{sec:magic_spectral_study}, the data is fitted with a power-law model between 100\,GeV and 2\,TeV. The best fit parameters can be found in Table~\ref{tab:MAGIC_spectral_param}. We plot the spectra obtained during the VHE flare period in August 2019, during the ``soft X-ray flare'' \& ``hard low X-ray state'' as well as during the deep exposures (on July 23\textsuperscript{rd} 2020 and August 6\textsuperscript{th} 2021). They are shown in Fig.~\ref{magic_seds_flares},~\ref{magic_seds_special_days} and ~\ref{magic_seds_deep_exposures}, respectively. For comparison purposes, we also show the average spectra from 2020 and 2021. For each fit the ``butterfly'' represents the 1-$\sigma$ uncertainty band.


\clearpage

\section{Modelling of the quiescent activity using a LPPL function for the electron distribution}
\label{low_state_modelling_lppl}

We also modelled the MAGIC, \textit{XMM-Newton}, \textit{NuSTAR}, and \textit{AstroSat} simultaneous nights using a LPPL function (instead of a broken power-law function) to describe the electron distribution within the ``variable'' region. The LPPL function, originally introduced by \citet{2006A&A...448..861M}, is defined in Eq.~\ref{LPPL_function}. The resulting parameter values are listed in Table~\ref{tab:ssc_parameters_deep_exposures_lppl}. We refer the reader to Sect.~\ref{low_state_model} for a detailed description of the respective parameters (as well as the procedure to constrain them).

\begin{table*}[h!]
\caption{\label{tab:ssc_parameters_deep_exposures_lppl} Parameters of the two-components SSC models obtained for the MAGIC, \textit{XMM-Newton}, \textit{NuSTAR}, and \textit{AstroSat} simultaneous observing nights using a LPPL function for the electron distribution in the ``variable'' region.}
\centering
\begin{tabular}{l c c c @{\hskip 0.3in} c}     
\hline\hline
   & \multicolumn{2}{c}{Variable region} &  & Core region \\\cline{2-3} 
Parameters   & ``deep exposure 1'' & ``deep exposure 2'' & & \\
   & July 23\textsuperscript{rd} 2020 & August 6\textsuperscript{th} 2021 & & \\
\hline
\hline
$B'$ [$10^{-2}$\,G]  & 5.6 & 6.3 & & 5.0  \\
$R'$ [$10^{16}$\,cm] & 2 & 2 & & 10 \\
$\delta$ & 10 & 10 & & 10\\
$U'_{\rm e}$ [$10^{-3}$\,erg cm$^{-3}$] & 6.9 & 5.0 & & 0.1\\
$s$ & 2.60 & 2.55 & & 2.15\\
$r$ & 1.00 & 2.17 & & --\\
$\gamma'_{\rm min}$ & $2\times10^{3}$ & $2\times10^{3}$ & & 10\\
$\gamma'_{\rm c}$ & $2.86\times10^{5}$ & $2.50\times10^{5}$ & & --\\
$\gamma'_{\rm max}$ & $8\times10^{6}$ & $8\times10^{6}$ & & $9.3\times10^{4}$\\

\hline
\end{tabular}
\tablefoot{See text in Sect.~\ref{low_state_model} for the description of each parameter.} 
\end{table*}

\end{appendix}

\end{document}